\documentclass[runningheads,a4page]{llncs}
\usepackage{wrapfig}
\usepackage{paralist}
\usepackage{mathrsfs}
\usepackage{amssymb}
\usepackage{amsmath}
\usepackage{times}
\usepackage{xcolor}
\usepackage[all]{xy}
\usepackage{tikz}
\usetikzlibrary{arrows,automata}
\usepackage{cases}
\usepackage{xparse,ifthen}
\usepackage{microtype}
\usepackage{algorithm2e}

\newboolean{useComments}
\setboolean{useComments}{false}
\ifthenelse{\boolean{useComments}}{%
  \newcommand{\todo}[1]{\textcolor{olive}{{#1}}}
  \newcommand{\ce}[1]{\textcolor{blue}{\texttt{ C E: {#1} :C E }}}
  \newcommand{\ls}[1]{\textcolor{teal}{\texttt{ L S: {#1} :S L }}}
  \newcommand{\hh}[1]{{\color{red}\texttt{ H H: #1 :H H }}}
  \newcommand{\js}[1]{{\color{orange}\texttt{ J S: #1 :J S }}}
  \newcommand{\lz}[1]{{\color{magenta}\texttt{ L Z: #1 :L Z }}}
}{%
  \newcommand{\todo}[1]{}
  \newcommand{\ce}[1]{}
  \newcommand{\ls}[1]{}
  \newcommand{\hh}[1]{}
  \newcommand{\js}[1]{}
  \newcommand{\lz}[1]{}
}
\newcommand{\initialState}{\bar{s}}

\newcommand{\TRAN}[1]{\xrightarrow[]{#1}}
\newcommand{\TRANsingle}[1]{\stackrel{#1}{\hookrightarrow}}
\newcommand{\WTRAN}[1]{\overset{#1}{\Longrightarrow}}
\newcommand{\TRANP}[1]{\xrightarrow[]{#1}_{\text{c}}}
\newcommand{\TRANPsingle}[1]{\TRANsingle{#1}_{\text{c}}}
\newcommand{\WTRANP}[1]{\overset{#1}{\Longrightarrow_{\text{c}}}}

\newcommand{\MC}[1]{\mathcal{#1}}

\newcommand{\ACT}{\mathit{Act}}
\newcommand{\ACTAU}{\mathit{Act}_{\tau}}

\newcommand{\RPLUS}{\mathbb{R}_{> 0}}

\newcommand{\STABLE}[1]{#1\!\downarrow}

\newcommand{\SUPP}{\mathit{Supp}}
\newcommand{\MI}[1]{\mathit{#1}}

\newcommand{\DIRAC}[1]{\delta_{#1}}

\newcommand{\PTRAN}[1]{\begin{tikzpicture}[->,>=stealth,auto,node distance=6pt,semithick]
	\tikzstyle{blackdot}=[circle,fill=black,minimum size=3pt,inner sep=0pt]
	\tikzstyle{stateNframe}=[draw=none,minimum size=1pt]	
    \node[stateNframe](n1){};
    \node[blackdot](n2)[right of=n1]{};
    \node[stateNframe](n3)[right of=n2]{};
    \path (n1.west) edge node {$\scriptstyle #1$~~}(n3.east);
\end{tikzpicture}
}

\newcommand{\MTRAN}[1]{
\begin{tikzpicture}[->,>=stealth,auto,node distance=6pt,semithick]
	\tikzstyle{stateNframe}=[draw=none,minimum size=1pt]	
    \node[stateNframe](n1){};
    \node[stateNframe](n3)[right of=n2]{};
    \path (n1.west) edge node {$\scriptstyle #1$~~}(n3.east)
    		(n1.west) edge node {}([xshift=-3pt]n3.east)
    		 (n1.west) edge node {}([xshift=-6pt]n3.east);
\end{tikzpicture}
}
\newcommand{\DIST}{\mathit{Dist}}
\newcommand{\ADIST}{\mathit{ADist}}

\newcommand{\ABS}[1]{|#1|\ }
\newcommand{\RATE}{\mathit{rate}}
\newcommand{\APAR}[3][A]{{#2}\parallel_{#1}{#3}}
\newcommand{\MA}{\text{\sf{MA}}}
\newcommand{\stateSet}{S}
\newcommand{\aMA}[1][]{(\stateSet#1,\ACTAU,\PTRAN{}#1,\MTRAN{}#1,\initialState#1)}

\newcommand{\DTIME}{+}

\newcommand{\M}{M}
\newcommand{\ML}{\MC{M}}

\newcommand{\EWBS}{^\bullet\!\!\!\approx}
\newcommand{\LWBS}{\approx\!\!\!^\bullet}

\newcommand{\PA}{\text{\sf PA}}
\newcommand{\IMC}{\text{\sf IMC}}

\newcommand{\TC}[1]{\overrightarrow{#1}}
\newcommand{\alphaup}{\theta}
\newcommand{\RGEZ}{\mathbb{R}_{\ge 0}}
\newcommand{\LAST}[1][\pi]{\mathit{last}(#1)}
\newcommand{\TTIME}{\MC{T}}
\newcommand{\ACTAUR}{\mathit{Act}_{\tau,r}}
\newcommand{\PATHS}{\MI{Paths}}
\newcommand{\SCH}{\xi}
\newcommand{\STEP}{\MI{Steps}}
\newcommand{\PR}{\MI{Pr}}

\newcommand{\MF}[1]{\mathfrak{#1}}
\newcommand{\EACT}{\MI{EA}}
\newcommand{\PROJ}[2]{[#1]_{#2}}
\newcommand{\TREQUIV}[1][]{\equiv_{#1}}
\DeclareMathAlphabet{\mathpzc}{OT1}{pzc}{m}{it}
\newcommand{\schedClass}{\mathpzc S}
\newcommand{\schedClassPartialInformation}{{{\mathpzc S}_P}}
\newcommand{\schedClassDistributedSchedulers}{{{\mathpzc S}_D}}
\newcommand{\TRACE}{\varsigma}

\newcommand{\CTMDP}{\textsf{CTMDP}}
\newcommand{\EHZ}{\approx
}

\newcommand{\distSetTC}{\TC{\mathcal{D}}}

\newcommand{\distHistoryU}{\mathcal{U}}
\newcommand{\distHistoryV}{\mathcal{V}}
\newcommand{\scheTran}[2]{\xrightarrow[]{#1}_{#2}}
\newcommand{\scheWTran}[2]{\overset{#1}{\Longrightarrow}_{#2}}
\def\dcup{\overset{.}{\cup}}
\newcommand{\traceCongruence}[1][\schedClass]{\equiv^c_{#1}}

\title{Late Weak Bisimilarity for Markov Automata}

\titlerunning{Late Weak Bisimilarity for MAs}
\authorrunning{Christian Eisentraut et al.}

\author{
Christian Eisentraut\inst{1}
\and
Jens Chr. Godskesen\inst{2}
\and \\
Holger Hermanns\inst{1}
\and
Lei Song\inst{3,1}
\and
Lijun Zhang\inst{4,1}
}

\institute
{
 \inst{}
  Saarland University, Germany
  \and
  \inst{}
  IT University of Copenhagen, Denmark
 \and
  \inst{}
  Max-Planck-Institut f\"{u}r Informatik, Germany
   \and
  \inst{} 
  Institute of Software, Chinese Academy of Sciences, China   
}

\begin{document}

\maketitle
\begin{abstract}
  Weak bisimilarity is a distribution-based equivalence notion
  for Markov automata. It has gained some popularity as the
  coarsest reasonable behavioural equivalence on Markov
  automata. This paper studies a strictly coarser notion: Late weak
  bisimilarity enjoys valuable properties if restricting to important
  subclasses of schedulers: Trace distribution equivalence is implied
  for partial information schedulers, and compositionality is preserved 
  by distributed schedulers. The intersection of the two scheduler classes thus
  spans a coarser and still reasonable compositional theory of Markov
  automata.
\end{abstract}
\sloppy
\section{Introduction}\label{sec:introduction}
Compositional theories have become a foundation for developing
effective techniques for analysing stochastic systems. Their potential
ranges from compositional minimization
\cite{garavel-96,BodeHHJPPRWB09,BoudaliCS10} approaches to component
based verification \cite{KwiatkowskaNPQ10,FHKP11}.

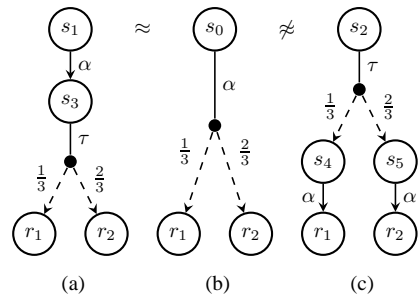
\begin{wrapfigure}{r}{0.5\textwidth}
  \centering
 \begin{tikzpicture}[->,>=stealth,auto,node distance=1.2cm,semithick,scale=0.8,every node/.style={scale=0.8}]
	\tikzstyle{blackdot}=[circle,fill=black,minimum size=6pt,inner sep=0pt]
	\tikzstyle{state}=[minimum size=0pt,circle,draw,thick]
	\tikzstyle{stateNframe}=[minimum size=0pt]	
	\node[state](r11){$r_1$};
	\node[state](r21)[right of=r11]{$r_2$};
	\node[state](r12)[right of=r21]{$r_1$};
	\node[state](r22)[right of=r12]{$r_2$};
	\node[state](r13)[right of=r22]{$r_1$};
	\node[state](r23)[right of=r13]{$r_2$};
	\node[state](s4)[above of=r13]{$s_4$};
	\node[state](s5)[above of=r23]{$s_5$};	
	\node[blackdot](d1)[above of=r11,xshift=0.6cm]{};
	\node[blackdot](d2)[above of=r12,xshift=0.6cm,yshift=0.6cm]{};
	\node[blackdot](d3)[above of=s4,xshift=0.6cm]{};
	\node[state](s3)[above of=d1,yshift=-0.2cm]{$s_3$};
	\node[state](s1)[above of=s3]{$s_1$};
	\node[stateNframe](f1)[right of=s1]{$\EHZ$};	
	\node[state](s0)[right of=f1]{$s_0$};
	\node[stateNframe](f2)[right of=s0]{$\not\EHZ$};	
	\node[state](s2)[right of=f2]{$s_2$};
	\node[stateNframe](a)[below right of=r11,xshift=-0.2cm,yshift=0cm]{(a)};	
	\node[stateNframe](b)[below right of=r12,xshift=-0.2cm,yshift=0cm]{(b)};
	\node[stateNframe](c)[below right of=r13,xshift=-0.2cm,yshift=0cm]{(c)};	
	\path (s1) edge								node {$\alpha$} (s3)
			  (s3) edge[-]							node {$\tau$} (d1)
			  (d1) edge[dashed]				node[left,yshift=6pt] {$\frac{1}{3}$} (r11)
			  		 edge[dashed]					node[yshift=-4pt] {$\frac{2}{3}$} (r21)
			  (s0) edge[-]							node {$\alpha$} (d2)			  
			  (d2) edge[dashed]				node[left, yshift=10pt] {$\frac{1}{3}$} (r12)
			  		 edge[dashed]					node {$\frac{2}{3}$} (r22)
			  (s2) edge[-]							node {$\tau$} (d3)
			  (d3) edge[dashed]				node[left,yshift=6pt] {$\frac{1}{3}$} (s4)
			  		 edge[dashed]					node[yshift=-4pt] {$\frac{2}{3}$} (s5)
			  (s4) edge								node[left] {$\alpha$} (r13)
			  (s5) edge								node {$\alpha$} (r23);
\end{tikzpicture}
  \caption{\label{fig:bisimulation ex}Examples of Markov automata.}
\vspace{-1.5em}
\end{wrapfigure}
Markov Automata (\MA{}s) are a compositional behavioural model for continuous
time stochastic and nondeterministic systems~\cite{Eisentraut2010CCS,eisentraut2010probabilistic}
subsuming Interactive Markov Chains (\IMC{}s)~\cite{HermannsIMC2002} and
Probabilistic Automata (\PA{}s)~\cite{Segala-thesis}. Markov automata weak probabilistic
bisimilarity has been introduced as an elegant and powerful way of
abstracting from internal computation cascades.  It is a conservative
extension of \IMC{}s weak bisimilarity, and also extends weak
probabilistic bisimilarity on \PA{}s. But different from standard
bisimulation notions, Markov automata weak bisimulations are defined
as relations on subprobability distributions instead of states. This
enables us to equate automata such as the ones on the left in
Fig.~\ref{fig:bisimulation ex}, but not the ones on the right, where
$\EHZ$ denotes the weak bisimilarity defined in~\cite{eisentraut2010probabilistic}.

An alternative formulation of \MA{} weak probabilistic bisimilarity
has later been coined \cite{dengsemantics} that, despite slight
differences in the setup, coincides with the original. As shown there,
weak probabilistic bisimilarity on \MA{}s can be considered as the
\emph{coarsest} equivalence relation preserving observable behaviour
and enjoying a congruence property with respect to parallel
composition. More precisely, it is shown to be the coarsest
reduction-closed barbed congruence~\cite{HondaT91}.

However, the relation discriminates automata, which one might 
intuitively expect to behave equivalent for every reasonable observer,
for instance the states $s_0$ and $s_2$ on the right of
Fig.~\ref{fig:bisimulation ex}.  We illustrate this with the following
example, inspired by \cite{Segala-thesis,Giro2007QMC}.

\begin{figure}[!t]
\begin{tabular}{|c|c|}
\hline
Tossing 1 & Tossing 2\\
 \begin{minipage}{0.45\linewidth}
    \begin{algorithm}[H]
     print(``I am going to toss")\;
    $r = \mathit{rand}()$\;
    \eIf{$r\ge \frac{1}{2}$}{
      print(``head")\;
     }{
      print(``tail")\;
    }
  \end{algorithm}
  \end{minipage} &
  \begin{minipage}{0.45\linewidth}
   \begin{algorithm}[H]
 $r = \mathit{rand}()$\;
 \eIf{$r\ge \frac{1}{2}$}{
	 print(``I am going to toss")\;   
   print(``head")\;
   }{
   print(``I am going to toss")\;
  print(``tail")\;
  }
\end{algorithm}
\end{minipage}\\
\hline
\end{tabular}
\caption{Two pieces of program used to simulate coin tossing.}\label{fig:program}
\vspace{-2em}
\end{figure}

\begin{wrapfigure}{r}{0.6\textwidth}
  \centering
  \begin{tikzpicture}[->,>=stealth,auto,node distance=1.7cm,semithick,scale=0.7, every node/.style={scale=0.7}]
	\tikzstyle{blackdot}=[circle,fill=black,minimum size=6pt,inner sep=0pt]
	\tikzstyle{state}=[minimum size=0pt,circle,draw,thick]
	\tikzstyle{stateNframe}=[minimum size=0pt]	
	\node[state](s31){$s_3$};
	\node[state](s41)[right of=s31]{$s_4$};
	\node[state](s32)[right of=s41]{$s_3$};
	\node[state](s42)[right of=s32]{$s_4$};
	\node[state](r5)[right of=s42]{$r_5$};
	\node[state](r6)[right of=r5]{$r_6$};
	\node[state](s11)[above of=s31]{$s_1$};
	\node[state](s21)[above of=s41]{$s_2$};
	\node[state](s12)[above of=s32]{$s_1$};
	\node[state](s22)[above of=s42]{$s_2$};
	\node[state](s5)[above of=s12]{$s_5$};
	\node[state](s6)[above of=s22]{$s_6$};
	\node[state](r3)[above of=r5]{$r_3$};	
	\node[state](r4)[above of=r6]{$r_4$};
	\node[state](r1)[above of=r3]{$r_1$};	
	\node[state](r2)[above of=r4]{$r_2$};
	\node[stateNframe](r00)[above of=r1]{};
	\node[state](r0)[right of=r00,xshift=-0.9cm]{$r_0$};
	\node[blackdot](d1)[above of=s11,xshift=0.9cm]{};  
	\node[state](s0)[above of=d1]{$s_0$};
	\node[state,scale=0.9](s0')[above of=s5]{$s'_0$};
	\node[blackdot](d2)[right of=s0',xshift=-0.9cm]{};
	\node[stateNframe](a)[below right of=s31,xshift=-10pt,yshift=0.5cm]{(a)};
	\node[stateNframe](b)[below right of=s32,xshift=-10pt,yshift=0.5cm]{(b)};
	\node[stateNframe](c)[below right of=r5,xshift=-10pt,yshift=0.5cm]{(c)};
	\path (s0) edge[-]							node {$i$} (d1)
			  (d1) edge[dashed]				node[left] {$\frac{1}{2}$} (s11)
			  		  edge[dashed] 				node[yshift=-9pt] {$\frac{1}{2}$} (s21)
			  (s0') edge[-]							node {$\tau$} (d2)
			  (d2) edge[dashed]				node[left] {$\frac{1}{2}$} (s5)
			  		  edge[dashed]				node[yshift=-9pt] {$\frac{1}{2}$} (s6)
			  (s5) edge								node[left] {$i$} (s12)
			  (s6) edge								node {$i$} (s22)
			  (s11) edge								node[left] {$h$} (s31)
			  (s21) edge								node {$t$} (s41)
			  (s12) edge								node[left] {$h$} (s32)
			  (s22) edge								node {$t$} (s42)
			  (r0) edge								node[left] {$i$} (r1)
			  		 edge								node[yshift=-8pt] {$i$} (r2)
			  (r1) edge								node[left] {$h$} (r3)
			  (r2) edge								node {$t$} (r4)
			  (r3) edge								node[left] {$\mathit{Suc}$} (r5)
			  (r4) edge								node {$\mathit{Suc}$} (r6);
  \end{tikzpicture}
  \caption{\label{fig:motivation}$s_0$ and $s'_0$ represent two different ways of tossing a coin and $r_0$ denotes the guessor.}
\vspace{-2em}
\end{wrapfigure}
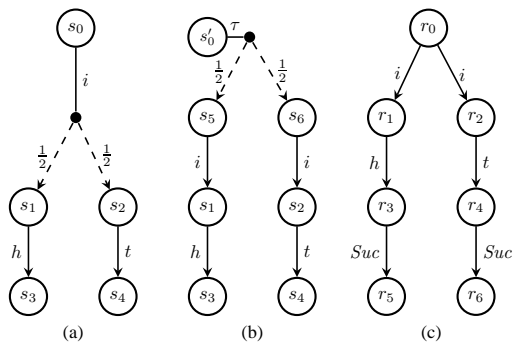

\begin{example}\label{ex:motivation}
  Refer to Fig~\ref{fig:program} for two pieces of program used to simulate
  coin tossing. We assume only ``print" is observable while others are non-observable.
  In ``Tossing 1", a sentence ``I am going to toss" is first printed to inform others who want
  to guess the tossing result. Then $r$ is assigned with a random number  in (0,1). If $r$
  is $\ge\frac{1}{2}$, ``head" is printed meaning that the coin tossing result is head,
  otherwise 	``tail" is printed. 
  Program	 ``Tossing 2" is slightly different. It first assigns $r$ with a random number
  in (0,1) as in ``Tossing 1". In case $r\ge\frac{1}{2}$, ``I am going to toss" is printed
  followed by the tossing result. Otherwise we obtain the tail of the coin.
  Intuitively, these two programs have no essential difference. However,
  when modelling them, we will obtain two different models, which are shown in 
  Fig.~\ref{fig:motivation}~(a) and~(b) respectively. Non-observable action is modelled by the
  internal action $\tau$  as in Fig.~\ref{fig:motivation} (b).
  In Fig.~\ref{fig:motivation}~(c) the guesser is modelled. While the
  tossing takes place (action $i$), he non-deterministically guesses
  the outcome, which he announces with the action $h$ or $t$, which
  stands for head or tail, respectively.

  The complete system is obtained by a
  parallel composition of the coin tosser automaton and the guesser
  automaton.  We use a \emph{CSP}-style parallel composition. Throughout
  our example, synchronization is enforced for actions in the set $A
  =\{i,h,t\}$.
   These actions synchronize
  with corresponding actions of the coin tosser. Thus, if the guess
  was right, the guesser finally performs the action $Suc$ to announce
  that he successfully guessed the outcome. \hfill\qed
\end{example}

In the example, the probability to see head or tail after a (fake)
coin toss is one half each, both for tosser (a) and (b). 
One would expect that hence the chance to guess correct is one half
for both tossers. However, $\APAR{s_0}{r_0}$
and $\APAR{s_0'}{r_0}$ are not weakly bisimilar, refer to Fig.~\ref{fig:execution}.
We will now show that the executions that distinguish the two
systems are actually caused by unrealistic schedulers, which cannot
appear in real world applications.  In Fig.~\ref{fig:execution}, we
color the execution of $\APAR{s_0'}{r_0}$ which is generated by a
scheduler that chooses transitions in a way such that
$\TRAN{\MI{Suc}}$ will be executed with probability 1. It is easy to
see that in contrast the probability that $\TRAN{\MI{Suc}}$ is
executed in $\APAR{s_0}{r_0}$ is at most 0.5, for every scheduler.

The intuitive reason why the scheduler for $\APAR{s_0'}{r_0}$ is too
powerful to be realistic is that it can base its decision which
transition to choose in state $r_0$ on the state the tosser has reached
by performing his internal probabilistic decision, namely either
state $s_5$ or $s_6$. If we consider the tosser and the guesser to be
independently running processes, this is not a realistic scheduler, as
then the guesser would need to see the internal state of the
tosser. However, no communication between guesser and tosser has
happened at this point in time, by which this information could have been
conveyed.
Thus, in distributed systems, where components only share the
information they gain through explicit communication via observable
actions, this behaviour is unrealistic. Thus, for practically relevant
models, weak bisimilarity for \MA{}s is still too coarse.  

In this paper, we present a novel notion of weak bisimilarity on
\MA{}s, called \emph{late weak bisimilarity}, that is coarser than the
existing notions of weak bisimilarity. It equates, for instance, the
two automata of Example~\ref{ex:motivation}, and all the ones
in~Fig.\ref{fig:bisimulation ex}. As weak bisimilarity is the coarsest
notion of equivalence that preserves observable behaviour and is
closed under parallel composition, late weak bisimilarity cannot
satisfy these properties in their entirety.  However, as we will show,
for a restricted class of schedulers, late weak bisimilarity preserves
observable behaviour, in the sense that trace distribution equivalence
$(i)$ is implied by late weak bisimilarity, and $(ii)$ is preserved in
the context of parallel composition. This also means that time-bounded
reachability properties are preserved with respect to parallel
composition.  The class of schedulers under which these properties are
satisfied is the intersection of two well-known scheduler classes,
namely partial information schedulers~\cite{de1999verification} and
distributed schedulers~\cite{Giro2007QMC}.  Both these classes have
been coined as principal means to exclude undesired or unrealistically
powerful schedulers. The co-inductive definition of late weak
bisimilarity we provide echoes these considerations on the automaton
level, thereby resulting in  a very coarse, yet reasonable, notion of
equality.

\section{Preliminaries}\label{sec:pre}
Let $\stateSet$ be a finite set of states ranged over by
$r,s,t,\ldots$. A \emph{distribution}\ce{is this the right
  term?}\ls{What do you mean?}\ce{usually distributions are defined to
  be full distributions. But i think it is too late now to make
  changes here in a consistent way. Maybe for the final version.}
is a function $\mu:\stateSet\to
[0,1]$ satisfying $\mu(\stateSet)=\sum_{s\in\stateSet}\mu(s)\le 1$. If
$\mu(\stateSet)=1$, it is called a \emph{full distribution}, otherwise
it is a \emph{sub-distribution}. Let $\ADIST(\stateSet)$ denote the
set of all (sub or full) distributions over $\stateSet$, ranged over
by $\mu,\nu,\gamma,\ldots$. Moreover, we use $\DIST(\stateSet)$ to
denote the set of all full distributions.  Define
$\SUPP(\mu)=\{s\mid\mu(s)>0\}$ as the support set of $\mu$. If
$\mu(s)=1$, then $\mu$ is called a \emph{Dirac} distribution, written
as $\DIRAC{s}$. Let $\ABS{\mu}=\mu(\stateSet)$ denote the size of the
distribution $\mu$. Given a real number $x$, $x\cdot\mu$ is the
distribution such that $(x\cdot\mu)(s)=x\cdot\mu(s)$ for each
$s\in\SUPP(\mu)$ if $x\cdot\ABS{\mu}\leq 1$, while $\mu-s$ is the
distribution such that $(\mu-s)(s)=0$ and $(\mu-s)(t)=\mu(t)$ with
$s\neq t$. Moreover, $\mu=\mu_1\DTIME\mu_2$ whenever $\mu(s)=\mu_1(s)
+ \mu_2(s)$ for each $s\in\stateSet$ 
and $\ABS{\mu}\leq 1$. We often write $\{s:\mu(s)\mid
s\in\SUPP(\mu)\}$ alternatively for a distribution $\mu$. For
instance, $\{s_1:0.4,s_2:0.6\}$ denotes a distribution $\mu$ such that
$\mu(s_1)=0.4$ and $\mu(s_2)=0.6$.

\begin{wrapfigure}{r}{0.7\textwidth}
  \begin{tikzpicture}[->,>=stealth,auto,node distance=2.5cm,semithick,scale=0.7, every node/.style={scale=0.7}]
	\tikzstyle{blackdot}=[circle,fill=black,minimum size=6pt,inner sep=0pt]
	\tikzstyle{state}=[rectangle,rounded corners=5,minimum size=15pt,draw,thick]
	\tikzstyle{stateNframe}=[minimum size=0pt]	
	\tikzstyle{bluestate}=[fill=none,rounded corners=5,draw=blue,text=blue,thick]
	\node[state](s0r0a){$\APAR[]{s_0}{r_0}$};
	\node[blackdot](d1)[above right of=s0r0a,yshift=-0.5cm,xshift=-0.3cm]{};
	\node[blackdot](d2)[below right of=s0r0a,yshift=0.5cm,xshift=-0.3cm]{};
	\node[state](s2r1a)[above right of=d1,yshift=-1cm]{$\APAR[]{s_2}{r_1}$};
	\node[state](s1r1a)[below right of=d1,yshift=1cm]{$\APAR[]{s_1}{r_1}$};
	\node[state](s3r3a)[right of=s1r1a]{$\APAR[]{s_3}{r_3}$};
	\node[state](s3r5a)[right of=s3r3a]{$\APAR[]{s_3}{r_5}$};
	\node[state](s2r2a)[above right of=d2,yshift=-1cm]{$\APAR[]{s_2}{r_2}$};
	\node[state](s1r2a)[below right of=d2,yshift=1cm]{$\APAR[]{s_1}{r_2}$};
	\node[state](s4r4a)[right of=s2r2a]{$\APAR[]{s_2}{r_4}$};
	\node[state](s4r6a)[right of=s4r4a]{$\APAR[]{s_4}{r_6}$};
	\node[stateNframe](a)[below of=s4r4a,yshift=0.5cm]{(a)};
	\node[bluestate](s0r0b)[below of=s0r0a,yshift=-3cm]{$\APAR[]{s'_0}{r_0}$};
	\node[blackdot,fill=blue](d3)[right of=s0r0b,xshift=-1cm]{};
	\node[bluestate](s6r0b)[below right of=d3,yshift=0.5cm]{$\APAR[]{s_6}{r_0}$};
	\node[bluestate](s5r0b)[above right of=d3,yshift=-0.5cm]{$\APAR[]{s_5}{r_0}$};
	\node[state](s1r2b)[above right of=s5r0b,yshift=-1cm,xshift=0.75cm]{$\APAR[]{s_1}{r_2}$};
	\node[bluestate](s2r2b)[above right of=s6r0b,yshift=-1cm,xshift=0.75cm]{$\APAR[]{s_2}{r_2}$};
	\node[state](s2r1b)[below right of=s6r0b,yshift=1cm,xshift=0.75cm]{$\APAR[]{s_2}{r_1}$};
	\node[bluestate](s4r4b)[right of=s2r2b]{$\APAR[]{s_4}{r_4}$};
	\node[bluestate](s4r6b)[right of=s4r4b]{$\APAR[]{s_4}{r_6}$};
	\node[bluestate](s1r1b)[below right of=s5r0b,yshift=1cm,xshift=0.75cm]{$\APAR[]{s_1}{r_1}$};
	\node[bluestate](s3r3b)[right of=s1r1b]{$\APAR[]{s_3}{r_3}$};
	\node[bluestate](s3r5b)[right of=s3r3b]{$\APAR[]{s_3}{r_5}$};
	\node[stateNframe](b)[below of=s2r1b,yshift=1.8cm]{(b)};
	\path (s0r0a)   edge[-]							node {$i$} (d1)
                        edge[-]							node[left] {$i$} (d2)
              (d1)      edge[dashed]						node[left,yshift=-5pt] {$\frac{1}{2}$} (s1r1a)
	                edge[dashed]						node {$\frac{1}{2}$} (s2r1a)	
	      (s1r1a)   edge							node {$h$} (s3r3a)
	      (s3r3a)   edge							node {$\mathit{Suc}$} (s3r5a)
	      (d2)      edge[dashed]						node {$\frac{1}{2}$} (s2r2a)
		        edge[dashed]						node[left,yshift=-5pt] {$\frac{1}{2}$} (s1r2a)
	      (s2r2a)   edge							node {$t$} (s4r4a)
	      (s4r4a)   edge							node {$\mathit{Suc}$} (s4r6a)
	      (s0r0b)   edge[-,draw=blue,text=blue]				node {$\tau$} (d3)		  
	      (d3)      edge[dashed,draw=blue,fill=blue,text=blue]		node {$\frac{1}{2}$} (s5r0b)		  
	                edge[dashed,draw=blue,fill=blue,text=blue]		node[left,yshift=-5pt] {$\frac{1}{2}$} (s6r0b)
	      (s5r0b)   edge				                        node {$i$} 
              (s1r2b)   edge[draw=blue,fill=blue,text=blue]			node[left,yshift=-5pt] {$i$} (s1r1b)
	      (s1r1b)	edge[draw=blue,fill=blue,text=blue]			node {$h$} (s3r3b)
	      (s3r3b)	edge[draw=blue,fill=blue,text=blue]			node {$\mathit{Suc}$} (s3r5b)
	      (s6r0b)   edge[draw=blue,fill=blue,text=blue]			node {$i$} (s2r2b)
	                edge							node[left,yshift=-5pt] {$i$} (s2r1b)
	      (s2r2b)	edge[draw=blue,fill=blue,text=blue]			node {$t$} (s4r4b)
	      (s4r4b)	edge[draw=blue,fill=blue,text=blue]			node {$\mathit{Suc}$} (s4r6b);			  	
  \end{tikzpicture}
  \caption{\label{fig:execution}
Executions of $\APAR{s_0}{r_0}$ and $\APAR{s'_0}{r_0}$ where $A$ is omitted.}
\vspace{-2em}
\end{wrapfigure}

\subsection{Markov Automata}
\begin{definition}
  \label{def:markov automata}
  An $\MA$ $\M$ is a tuple $\aMA{}$
  where $\initialState$ is the initial state, $\stateSet$ is a finite but
  non-empty set of states, $\ACTAU=\ACT \dcup \{\tau\}$ is a set of
  actions including the internal action $\tau$, $\PTRAN{}\ \subset
  \stateSet\times\ACTAU\times\DIST(\stateSet)$ is a finite set of
  probabilistic transitions, and $\MTRAN{}\ \subset\
  \stateSet\times\RPLUS\times\stateSet$ is a finite set of Markovian
  transitions.
\end{definition}


Let $\alpha,\beta,\gamma,\ldots$ range over the actions in $\ACTAU$
and $\lambda$ over the rates in $\RPLUS$. Moreover, let $\theta$ and
$\theta'$ range over $\ACTAUR=\ACTAU\cup\RGEZ$.
Let $\RATE(s,s')=\sum\{\lambda\mid (s,\lambda,s')\in\MTRAN{}\}$ (with empty
sum equal to $0$) denote the rate from $s$ to $s'$. We overload 
$\RATE$ to also denote the exit rate
of a state $s$ by writing
$\RATE(s)=\sum_{s'\in\stateSet}\RATE(s,s')$. A state $s\in\stateSet$ is
stable, written as $\STABLE{s}$, iff there does not exist $\mu$ such that 
$(s,\tau,\mu)\in\PTRAN{}$, similarly a
distribution $\mu$ is stable, written as $\STABLE{\mu}$, iff
$\STABLE{s}$ for each $s\in\SUPP(\mu)$.
For a stable state $s$, the sojourn time at $s$ is
exponentially distributed with rate equal to $\RATE(s)$, thus the
probability of leaving state $s$ within time interval $[a,b]$ is equal
to $e^{-\RATE(s)a}-e^{-\RATE(s)b}$. If more than one Markovian
transition is enabled from $s$, there is a race between them, and the
probability that the transition to state $s'$ is taken within $[a,b]$
is given by $(e^{-\RATE(s)a}-e^{-\RATE(s)b})\cdot
\frac{\RATE(s,s')}{\RATE(s)}$.
We write $s \TRAN{\theta} \mu$ if either
\begin{inparaenum}[\scriptsize$(i)$]
\item $\theta \in \ACTAU$  and $s \PTRAN{\theta} \mu$ or
\item $\theta \in \RPLUS$, $\STABLE{s}$, $\RATE(s)
  =\theta$, and for every state $s'$, $\mu(s') = \frac{\RATE(s, s')}{\RATE(s)}$, or
\item $\theta = 0$, $\mu = \DIRAC{s}$, $\RATE(s) = 0$ and
  $\STABLE{s}$.
\end{inparaenum}
This notation unifies immediate transitions~{\scriptsize$(i)$} and
timed transitions~{\scriptsize$(ii)$}.  As in
\cite{HermannsIMC2002,eisentraut2010probabilistic}, we make the
\emph{maximal progress assumption} and encoded it with this
notation. It says that if state $s$ is not stable, no Markovian
transitions can be executed.  Clause~{\scriptsize$(iii)$} generalizes
the implicit tangibility check of Clause~{\scriptsize$(ii)$} to states
without outgoing timed transitions. This generalization is needed to
encode a stability check in weak bisimilarity, which is inherited from
\IMC{}s, and necessary to achieve compositionality.

We remark that $\MA$s extend the well-known probabilistic automata
($\PA$s) \cite{Segala-thesis} and interactive Markov chains ($\IMC$s)
\cite{HermannsIMC2002}. Precisely, if $\MTRAN{}~=~\emptyset$, we
obtain $\PA$s. On the other side, if distributions are all Dirac,
i.e., $\PTRAN{}~\subset~\stateSet\times\ACTAU\times\DIRAC{\stateSet}$
with $\DIRAC{\stateSet}=\{\DIRAC{s}\mid s\in\stateSet\}$, we obtain
$\IMC$s.

\subsection{Schedulers}
We now recall some notations from \cite{Neuhausser2010TRP} defined
originally for $\CTMDP$s. Let throughout the paper $\ML$ denote the
$\MA$ $\aMA$. Finite
paths of $\ML$ are sequences like $\pi=s_0,\alphaup_0,t_0,\ldots,s_n$,
where $s_0=\initialState$ and $s_i\in\stateSet$ are states on the path
and $t_i\in\RGEZ$ is the sojourn time in state $s_i$. Recall that
$\alphaup_i$ is either an action in $\ACTAU$ or a Markovian action in
$\RGEZ$. Moreover $\alphaup_i\in\ACTAU$ implies $t_i=0$. The length of
$\pi$, denoted as $\ABS{\pi}$, is equal to the number of states on
$\pi$, and $\LAST=s_n$ is the last state on $\pi$. Let
$\TTIME(\pi)=\sum_{0\leq i< n}t_i$ be the total time spent on $\pi$,
$\pi[n]$ denote the $n$-th state in $\pi$, and $\pi[0..n]$ denote the
prefix of $\pi$ with length $n$. Let $\pi\circ(\alphaup,t,s)$ denote
a path obtained by extending $\pi$ with $(\alphaup,t,s)$. Let
$\Omega=\ACTAUR\times\RGEZ\times\stateSet$, then
$\PATHS^n(\ML)=\stateSet\times\Omega^n$ is the set of paths of $\ML$
with length $n$. Accordingly, let $\PATHS^{*}(\ML)$,
$\PATHS^{\omega}(\ML)$, and $\PATHS(\ML)$ denote the set of finite,
infinite, and all paths of $\ML$, respectively. In case $\ML$ is
replaced by a state $s$, they are constrained to paths starting from
$s$.  For simplicity we shall omit the script $\ML$ in the following
if it is clear from the context.  Define
$\MF{F}=\sigma(2^{\ACTAUR}\times\MF{B}\times2^{\stateSet})$ as the
$\sigma$-field over subsets of $\Omega$, where $\MF{B}$ is the Borel
$\sigma$-field over $\RGEZ$. According to standard measure theory,
$\MF{F}_{\PATHS^n}=\sigma(\{S_0\times\MI{tr}_0\times\ldots\times\MI{tr}_n\mid
S_0\in2^{\stateSet}\land\MI{tr}_i\in\MF{F}\})$ are measurable subsets
of $\PATHS^n$. Given $\Pi\in\MF{F}_{\PATHS^n}$, a \emph{cylinder} $C$
based on $\Pi$ can be defined as follows:
$C=\MI{Cyl}(\Pi)=\{\pi\in\PATHS^{\omega}\mid\pi[0..n]\in \Pi\}$. The
$\sigma$-field
$\MF{F}_{\PATHS^{\omega}}=\sigma(\cup^{\infty}_{n=0}\{\MI{Cyl}(\Pi)\mid
\Pi\in\MF{F}_{\PATHS^n}\})$ contains all the cylinders.

As usual, we need to resolve non-determinism before we can
define a probability measure for paths of a given $\MA$. This is
done by introducing schedulers. Intuitively, a scheduler will decide
how to resolve non-deterministic choices probabilistically based on
some prior information like the states visited, the elapsed time and
so on. Let $\STEP(s)=\{(\alphaup,\mu)\mid s\TRAN{\alphaup}\mu\}$
denote the set of transitions enabled at $s$. Below follows the formal
definition of schedulers.
\begin{definition}
\label{def:scheduler}
A scheduler $\SCH$ of $\ML$ is a function
$\PATHS^{*}\times \ACTAUR\times\DIST(\stateSet) \mapsto [0,1]$ with
$\SCH(\pi,\cdot,\cdot)\in\DIST(\STEP(\LAST))$ for all $\pi\in\PATHS^{*}$ and where
$\SCH(\cdot,\alphaup,\mu):\PATHS^{*}\mapsto[0,1]$ are measurable for all
$(\alphaup,\mu)\in 2^{\ACTAUR\times\DIST(\stateSet)}$.
\end{definition}

Given a scheduler $\SCH$ of $\ML$, we can now define a unique
probability measure
$\PR_{\initialState}^{\SCH}:\MF{F}_{\PATHS^{\omega}}\mapsto[0,1]$ on
$(\PATHS^{\omega},\MF{F}_{\PATHS^{\omega}})$.  The measure
$\PR_{\initialState}^{\SCH}$ is defined inductively as follows: if
$\Pi\in\MF{F}_{\PATHS^0}$, then
$\PR_{\initialState}^{\SCH}=I_{\Pi}(\initialState)$, where $I$ is the
characteristic function of set $\Pi$. If
$\Pi\in\MF{F}_{\PATHS^{n+1}}$, then $\PR_{\initialState}^{\SCH}(\Pi)=$
$${\small
\int_{\pi\in\PATHS^n}\PR_{\initialState}^{\SCH}(\pi)\left(\sum_{(\alphaup,\mu)\in\STEP(\LAST)}\SCH(\pi,\alphaup,\mu)\eta(\alphaup,t)\sum_{s'\in\stateSet}I_{\Pi}(\pi\circ(\alphaup,t,s'))\mu(s')\right)d\pi}$$
where $\eta(\alphaup,t)=1$ if $\alphaup\in\ACTAU\land t=0$,
$\eta(\alphaup,t)=0$ if $\alphaup\in\ACTAU\land t\neq 0$, and
$\eta(\alphaup,t)=\lambda e^{-\lambda t}$ if $\alphaup=\lambda$.
Intuitively, the value of $\PR_{\initialState}^{\SCH}(\Pi)$ is equal
to the sum of $\PR_{\initialState}^{\SCH}(\pi\circ(\alphaup,t,s'))$
for each $\pi\circ(\alphaup,t,s')\in\Pi$, where the value of
$\PR_{\initialState}^{\SCH}(\pi\circ(\alphaup,t,s'))$ is inductively
determined by the product of four probabilities:
\begin{inparaenum}[\scriptsize$(i)$]
\item the probability of the prefix of $\pi$ with length $n$, given $\SCH$: $\PR_{\initialState}^{\SCH}(\pi)$;
\item the probability of $(\alphaup,\mu)\in\STEP(\LAST)$ being chosen by $\SCH$, given $\pi$: $\SCH(\pi,\alphaup,\mu)$;
\item the probability of staying at state $\LAST$ for $t$ time units:
  $\eta(\alphaup,t)$;
\item the probability of $s'$ in $\mu$: $\mu(s')$.
\end{inparaenum}
The characteristic function  $I_{\Pi}(\pi\circ(\alphaup,t,s'))$ guarantees 
that we only count paths in $\Pi$.
Function $\eta(\alphaup,t)$ is the probability of staying at state $\LAST$
for $t$ time units before performing the transition labelled with
$\alphaup$. Therefore if $\alphaup=\lambda$, it is equal to $\lambda
e^{-\lambda t}$. If $\alphaup\in\ACTAU$, it must be case that $t=0$,
otherwise we let $\eta(\alphaup,t)=0$ to ignore impossible paths.
\section{Weak Bisimilarities for Markov Automata}\label{sec:weak bisimilarities}
In this section, we first introduce early weak bisimulation, which is a
variant of weak bisimulation defined in~\cite{dengsemantics}, and then define
late weak bisimulation, which is strictly coarser than early weak bisimulation.  

We first introduce a standard weak transition relation needed in the
definitions of bisimulation that allows to abstract from internal
actions.  Intuitively, $s\WTRAN{\theta}\mu$ denotes that a
distribution $\mu$ is reached from $s$ by a $\theta$-transition, which
may be preceded and followed by an arbitrary sequence of internal
transitions.  Formally, we define them as
derivations~\cite{DengGHM09,dengsemantics} for $\MA$s. In the
following, let $\mu\TRAN{\theta}\mu'$ iff there exists a transition
$s\TRAN{\theta}\mu_{s}$ for each $s\in\SUPP(\mu)$ such that
$\mu'=\sum_{s\in\SUPP(\mu)}\mu(s)\cdot\mu_s$. Then, $s\WTRAN{\tau}\mu$
iff there exists $ \DIRAC{s} = \mu_0^{\to} + \mu_0^{\times},\
\mu_0^{\to} \TRAN{\tau} \mu_1^{\to} +\mu_1^{\times},\ \mu_1^{\to}
\TRAN{\tau} \mu_2^{\to} + \mu_2^{\times},\ \ldots, $ where
$\mu=\sum_{i\ge 0}\mu_i^{\times}$.  We write $s\WTRAN{\theta}\mu$ iff
there exists $s\WTRAN{\tau}\TRAN{\theta}\WTRAN{\tau}\mu$.
Given a transition relation $\mathord{\rightsquigarrow} \subseteq
\stateSet \times \ACTAU \times \DIST({\stateSet})$, we let 
 $s\stackrel{\theta}{\rightsquigarrow}_c\mu$ iff there exists a
finite number of real numbers $w_i>0$, and transitions $s
\stackrel{\theta}{\rightsquigarrow} \mu_i$ such that $\sum_i w_i=1$,
and $\sum_i w_i\cdot\mu_i=\mu$. We call 
$\rightsquigarrow_c$ \emph{combined transitions} (of $\rightsquigarrow$).
In general, we lift a transition relation $\rightsquigarrow \subseteq
\stateSet \times \ACTAU \times \DIST({\stateSet})$ over states to a
transition relation $ \DIST({\stateSet}) \times \ACTAU \times
\DIST({\stateSet})$ over distributions by letting
$\mu\stackrel{\theta}{\rightsquigarrow}\mu'$ iff there exists a transition
$s\stackrel{\theta}{\rightsquigarrow}\mu_{s}$ for each $s\in\SUPP(\mu)$ such that
$\mu'=\sum_{s\in\SUPP(\mu)}\mu(s)\cdot\mu_s$.

\begin{definition}
\label{def:weak bisimulation early}
  A relation
  $\MC{R}\subseteq\DIST(\stateSet)\times\DIST(\stateSet)$ is an early weak
  bisimulation over $\ML$ iff $\mu~\MC{R}~\nu$ implies:
  \begin{inparaenum}[\scriptsize$(i)$]
  \item\label{item:early 1} whenever $\mu\TRANP{\theta}\mu'$, there exists a
    $\nu\WTRANP{\theta}\nu'$ such that $\mu'~\MC{R}~\nu'$;
  \item\label{item:early 2} whenever $\mu=\sum_{0\leq i\leq n}p_i\cdot\mu_i$, there exists
    $\nu\WTRANP{\tau}\sum_{0\leq i\leq n}p_i\cdot\nu_i$ such that
    $\mu_i~\MC{R}~\nu_i$ for each $0\leq i\leq n$ where $\sum_{0\leq
      i\leq n}p_i=1$;
  \item\label{item:early 3} symmetrically for $\nu$.
  \end{inparaenum}
   We say that $\mu$ and $\nu$ are early weak bisimilar, written as
  $\mu~\EWBS~\nu$, iff there exists an early weak bisimulation
  $\MC{R}$ such that $\mu~\MC{R}~\nu$. Moreover $s~\EWBS~r$ iff
  $\DIRAC{s}~\EWBS~\DIRAC{r}$.
\end{definition}
Clause~$(\ref{item:early 1})$ is standard. Clause~$(\ref{item:early 2})$ says that no matter how we split $\mu$,
there always exists a splitting of $\nu$ probably after internal
transitions to simulate the splitting of $\mu$.
Definition~\ref{def:weak bisimulation early} is slightly different
from Definition~5 in~\cite{dengsemantics}, where Clause~$(\ref{item:early 2})$ is missing and
Clause~$(\ref{item:early 1})$ is replaced by: whenever $\mu\WTRANP{\theta}\sum_{0\le i\le n}p_i\cdot\mu_i$,
there exists $\nu\WTRANP{\theta}\sum_{0\le i\le n}p_i\cdot\nu_i$ such that
$\mu_i~\MC{R}~\nu_i$ for each $0\le i\le n$. Essentially,
this condition subsumes Clause~$(\ref{item:early 2})$, since $\mu=\sum_{0\le i\le n}p_i\cdot\mu_i$
implies $\mu\WTRANP{\tau}\sum_{0\le i\le n}p_i\cdot\mu_i$.
As we shall prove later, both definitions induce the same equivalence
relation on Markov automata.
Clause~$(\ref{item:early 2})$ in Definition~\ref{def:weak bisimulation
  early} is, in fact, the cause why this relation is unrealistically
strong for scenarios as those discussed in
Example~\ref{ex:motivation}. The reason is that in order to establish
a bisimulation, \emph{every} splitting of $\mu$ into subdistributions
must be matched by $\nu$ (possibly after some internal transitions).
This also includes splittings into Dirac distributions. Intuitively,
this means that still the individual behaviour of each single state in
$\SUPP(\mu)$ must be matched.  In our scenarios, however, we
want to focus on the behaviour of distributions over states and not their
 individual supporting states. We will correct this in our definition
 of late weak bisimulation later. We still need to introduce a few notions beforehand.

\begin{definition}
  A distribution $\mu$ is transition consistent, written as
  $\TC{\mu}$, if for any $s\in\SUPP(\mu)$ and $\theta\not\in\{\tau,0\}$,
  $s\WTRAN{\theta}\gamma$ for some $\gamma$ implies
  $\mu\WTRAN{\theta}\gamma'$ for some $\gamma'$.
\end{definition}
Intuitively, if a distribution is transition consistent, all states in
its support have the same set of enabled visible actions.  When a
distribution is transition consistent, then $\mu \WTRAN{\theta}$
whenever there is a a state $s\in\SUPP(\mu)$ with
$s\WTRAN{\theta}$. This also means that when a distribution is
\emph{not} transition consistent, then there may be a weak $\theta$
transition that a certain state in the support can perform but the
distribution cannot. We then say that this state is \emph{blocked}
from taking this transition. When we adopt the notion of blocked
states accordingly for non-weak transition relations, also $\tau$
transitions can be blocked.

We now introduce $\TRANsingle{}$, an alternative lifting of
transitions of states to transitions of distributions that differs
from the standard definition used
in~\cite{eisentraut2010probabilistic,dengsemantics}. There, a
distribution is able to perform a transition labelled with $\theta$
\emph{if and only if} all the states in its support can perform
transitions with the very same label. In contrast, the transition relation
$\TRANsingle{}$ behaves like a weak transition, where every state in
the support of $\mu$ may at most perform one transition.

\begin{definition}
\label{def:lift transition to distributions}
  $\mu\TRANsingle{\theta}\mu'$
     iff either
  \begin{inparaenum}[\scriptsize$(i)$]
  \item\label{item:lift 1} for each $s\in\SUPP(\mu)$ there exists $s\TRAN{\theta}\mu_s$
    such that $\mu'=\sum_{s\in\SUPP(\mu)}\mu(s)\cdot\mu_s$ or,
  \item\label{item:lift 2} $\theta=\tau$ and there exists $s\in\SUPP(\mu)$ and
    $s\TRAN{\theta}\mu_s$ such that $\mu'=(\mu-s) + \mu(s)\cdot\mu_s$.
  \end{inparaenum}
\end{definition}
In the definition of late weak bisimulation, this extension will be
used to prevent $\tau$ transitions of states from being blocked.
Below follows an example: 
\begin{example}\label{ex:distribution transition}
  Let $\mu=\{s_1:0.4,s_2:0.6\}$ such that $s_1\TRAN{\tau}\DIRAC{s'_1}\TRAN{\alpha}\mu_1$,
  $s_1\TRAN{\beta}\mu_2$, $s_2\TRAN{\alpha}\mu_3$, and
  $s_2\TRAN{\beta}\mu_4$, where $\alpha\neq\beta$ are visible
  actions. According to Clause~$(\ref{item:lift 1})$ of Definition~\ref{def:lift transition
    to distributions}, we will have
  $\mu\TRANsingle{\beta}(0.4\cdot\mu_2 + 0.6\cdot\mu_4)$.   Without
 Clause~$(\ref{item:lift 2})$, this would be the only transition of $\mu$,
  since the $\tau$ transition of $s_1$ and the $\alpha$ transition of
  $s_2$ will be blocked by each other,  as $s_1$ and $s_2$ cannot
  perform transitions with labels $\tau$ and $\alpha$ at the same time.
  
  Note that the $\alpha$ transition is blocked by the $\tau$
  transition of $s_1$, so according to Clause~$(\ref{item:lift 2})$ of
  Definition~\ref{def:lift transition to distributions}, we in
  addition have $\mu\TRANsingle{\tau}(0.4\cdot\DIRAC{s'_1} +
  0.6\cdot\DIRAC{s_2})\TRANsingle{\alpha}(0.4\cdot\mu_1+0.6\cdot\mu_3)$.
 Note that in Clause~{\scriptsize$(i)$} of
  Definition~\ref{def:weak bisimulation early}, $\TRAN{}$ can be
  replaced by $\TRANsingle{}$ without changing the resulting
  equivalence relation, as the same effect can be obtained by a
  suitable splitting in Clause~{\scriptsize$(ii)$}. In this example,
  we could let $\mu$ be split into
  $0.4\cdot\DIRAC{s_1}+0.6\cdot\DIRAC{s_2}$, such that no transition
  is blocked in the resulting distributions.
\end{example}

\begin{definition}
\label{def:weak bisimulation late} 
A relation
$\MC{R}\subseteq\DIST(\stateSet)\times\DIST(\stateSet)$ is a late weak
bisimulation over $\ML$ iff $\mu~\MC{R}~\nu$ implies:
  \begin{inparaenum}[\scriptsize$(i)$]
  \item\label{item:late 1} whenever $\mu\TRANPsingle{\theta}\mu'$, there exists a
$\nu\WTRANP{\theta}\nu'$ such that $\mu'~\MC{R}~\nu'$;
  \item\label{item:late 2} if not $\TC{\mu}$, then there exists $\mu=\sum_{0\leq i\leq
n}p_i\cdot\mu_i$ and $\nu\WTRANP{\tau}\sum_{0\leq i\leq
n}p_i\cdot\nu_i$ such that $\TC{\mu_i}$ and $\mu_i~\MC{R}~\nu_i$ for
each $0\leq i\leq n$ where $\sum_{0\leq i\leq n}p_i=1$;
\item\label{item:late 3} symmetrically for $\nu$.
  \end{inparaenum} 
We say that $\mu$ and $\nu$ are late weak bisimilar,
written as $\mu~\LWBS~\nu$, iff there exists a late weak bisimulation
$\MC{R}$ such that $\mu~\MC{R}~\nu$. Moreover $s~\LWBS~r$ iff
$\DIRAC{s}~\LWBS~\DIRAC{r}$.
\end{definition} 
\begin{wrapfigure}{r}{0.45\textwidth}
 \centering 
 \begin{tikzpicture}[->,>=stealth,auto,node distance=1.5cm,semithick,scale=0.7, every node/.style={scale=0.7}]
	\tikzstyle{blackdot}=[circle,fill=black,minimum size=6pt,inner sep=0pt]
	\tikzstyle{state}=[circle,minimum size=0pt,draw,thick]
	\tikzstyle{stateNframe}=[minimum size=0pt]	
	\tikzstyle{greenstate}=[fill=none,rounded corners=5,draw=green,text=green,thick]
	\node[state](r2a){$r_2$};
	\node[state](r3a)[right of=r2a]{$r_3$};
	\node[state](r1b1)[right of=r3a]{$r_1$};
	\node[state](r2b)[right of=r1b1,xshift=-0.5cm]{$r_2$};
	\node[state](r3b)[right of=r2b]{$r_3$};
	\node[state](r1b2)[right of=r3b,xshift=-0.5cm]{$r_1$};
	\node[blackdot](d1)[above of=r2a,xshift=1cm,yshift=1cm,label={[label distance=0pt]0:{$\mu$}}]{};
	\node[state](r1a)[above of=r2a,yshift=0.75cm]{$r_1$};
	\node[state](s0)[above of=d1,yshift=0.6cm]{$s_0$};
	\node[state](s2)[above of=r2b]{$s_2$};
	\node[state](s3)[above of=r3b]{$s_3$};
	\node[blackdot](d2)[above of=s2,xshift=0.75cm,label={[label distance=0pt]0:{$\nu$}}]{};
	\node[state](s1)[above of=d2]{$s_1$};
	\path (s0) edge 						node[left] {$\beta$} (r1a)
					edge[-]						node[right] {$\alpha$} (d1)
				(d1) edge[dashed]		node[left] {$\frac{1}{3}$} (r2a)
						edge[dashed]		node[right] {$\frac{2}{3}$} (r3a)
				(s1) edge[-]					node[left] {$\tau$} (d2)
				(d2) edge[dashed]      node[left] {$\frac{1}{3}$} (s2)
						edge[dashed]		node[right] {$\frac{2}{3}$} (s3)
				(s2) edge						node[left] {$\beta$} (r1b1)
					   edge						node[right] {$\alpha$} (r2b)
				(s3) edge						node[left] {$\alpha$} (r3b)
					   edge						node[right] {$\beta$} (r1b2);
\end{tikzpicture}
  \caption{\label{fig:pa}$s_0~\LWBS~r_0$.}
\vspace{-2em}
\end{wrapfigure}

In Clause~$(\ref{item:late 1})$, this definition differs from Definition~\ref{def:weak
  bisimulation early} by the use of $\TRANsingle{}$. It is
straightforward to show that $\TRANsingle{}$ can also be used in
Definition~\ref{def:weak bisimulation early} without changing the
resulting bisimilarity. However, in Definition~\ref{def:weak bisimulation
  late}, using $\TRAN{}$ instead of $\TRANsingle{}$ will lead to a
finer relation.  The key difference between Definition~\ref{def:weak
  bisimulation early} and \ref{def:weak bisimulation late}, however,
is Clause~$(\ref{item:late 2})$. In Definition~\ref{def:weak
  bisimulation early}, we require that for \emph{any} split of $\mu$ such
that $\mu=\sum_{0\leq i\leq n}p_i\cdot\mu_i$, there exists
$\nu\WTRANP{\tau}\sum_{0\leq i\leq n}p_i\cdot\nu_i$ with
$\mu_i~\MC{R}~\nu_i$ for each $i$, while in Definition~\ref{def:weak
  bisimulation late}, we require to split $\mu$ \emph{only if} it is not
transition consistent. We further require that the resulting
distributions $\mu_i$ are transition consistent. We do not require
this for the $\nu_i$. It can be shown, however, that $\TC{\mu_i}$ and
$\mu_i~\MC{R}~\nu_i$ implies $\TC{\nu_i}$.  These conditions ensure
that no states in the support of $\mu$ are blocked from executing
certain transitions \emph{for ever}.  
Clearly, if $\mu$ is already
transition consistent, we do not need to split $\mu$ further, since no
transition of states in $\SUPP(\mu)$ are blocked, and thus the
distribution transitions in Clause {\scriptsize$(i)$} suffice to
capture every visible behaviour.

\begin{remark} Essentially, in Definition~\ref{def:weak bisimulation
late} we keep all states with the same set of enabled actions
together. This is similar to the idea in~\cite{de1999verification},
where all states with the same enabled actions are
non-distinguishable from the outside. Once a distribution becomes
transition consistent, we will not try to split it anymore -- but
rather match the lifted transitions according to the first clause. 
\end{remark}

\begin{example}\label{ex:ex1} 
We will show that in Fig.~\ref{fig:motivation} , $s_0~\LWBS~s'_0$. Let
$\MC{R}=\{(\DIRAC{s_0},\DIRAC{s'_0}),(\DIRAC{s_0},\{s_5:0.5,s_6:0.5\})\}\cup\MI{ID}$
where $\MI{ID}$ is the identity relation. It is easy to show that
$\MC{R}$ is a late weak bisimulation. The only non-trivial case is
when $\DIRAC{s'_0}\TRAN{\tau}\{s_5:0.5,s_6:0.5\}$. But then $\DIRAC{s_0}$
can simulate it without performing any transition
i.e. $\DIRAC{s_0}\WTRAN{\tau}\DIRAC{s_0}$. Since
$(\DIRAC{s_0},\{s_5:0.5,s_6:0.5\})\in\MC{R}$, Clause~$(\ref{item:late 1})$ of
Definition~\ref{def:weak bisimulation late} is satisfied. Moreover
both $\DIRAC{s_0}$ and $\{s_5:0.5,s_6:0.5\}$ are transition
consistent, thus we do not need to split them any further.
Conversely, we can show that $\MC{R}$ is not an early weak
bisimulation. According to Clause~$(\ref{item:early 1})$ of Definition~\ref{def:weak
bisimulation early}, we require that for any split of
$\{s_5:0.5,s_6:0.5\}$, there must exist a matching split of
$\DIRAC{s_0}$, which cannot be established. For instance the split
$\{s_5:0.5,s_6:0.5\}=0.5\cdot\DIRAC{s_5}+0.5\cdot\DIRAC{s_6}$ cannot
be matched by any split of $\DIRAC{s_0}$.\hfill\qed
\end{example}

The following example shows that the transition consistency condition
of Definition~\ref{def:weak bisimulation late} is necessary to not
equate  states which should be distinguished.
\begin{example} Suppose there are two states $s_0$ and $r_0$ such that
  $s_0\TRAN{\tau}s_1$ and $r_0\TRAN{\tau}\{r_1:0.5,r_2:0.5\}$ where
  all of $s_1$, $r_1$, and $r_2$ have a transition to themselves with
  labels $\tau$, in addition, $r_1\TRAN{\alpha}r_1$ where
  $\alpha\neq\tau$. Let
  $\MC{R}=\{(\DIRAC{s_0},\DIRAC{r_0}),(\DIRAC{s_1},\{r_1:0.5,r_2:0.5\})\}$. If
  we dropped the transition consistency condition from
  Definition~\ref{def:weak bisimulation late}, we could show that
  $\MC{R}$ is a late weak bisimulation, and therefore $s_0~\LWBS~r_0$,
  because the distribution $\{r_1:0.5,r_2:0.5\}$ can only perform a
  $\tau$ transition to itself, while the $\alpha$ transition of $r_1$
  would then be blocked. However, $s_0$ and $r_0$ should be
  distinguished, because $r_0$ can reach $r_1$ with positive
  probability, which is a state able to perform a transition with
  visible label $\alpha$.  Note that as $\{r_1:0.5,r_2:0.5\}$ is not
  transition consistent, we should split it further according to
  Definition~\ref{def:weak bisimulation late}. Thus we can prove that
  $\MC{R}$ is not a late weak bisimulation i.e. $s_0~\not\LWBS~r_0$.
\end{example}

Since we treat Markovian transitions in the same way as non-Markovian
transitions, Definition~\ref{def:weak bisimulation late} also applies
for $\PA$s, a subset of $\MA$s without Markovian transitions:

\begin{example}\label{ex:pa} Let $s_0$ and $s_1$ be two states in
Fig.~\ref{fig:pa}, where we omit the transitions of $r_1$, $r_2$, and
$r_3$. Note that in $s_0$ the probabilistic transition is after the
$\alpha$ transition, while in $s_1$ the probabilistic transition is
before the $\alpha$ transition. It is routine to check that $s_0$ and
$s_1$ are late weak bisimilar, but not early weak bisimilar. Since
$\nu$ is transition consistent, and can be simulated by $\DIRAC{s_0}$
according to Definition~\ref{def:weak bisimulation late}. But for
early weak bisimulation, there exists a split
$\nu=\frac{1}{3}\cdot\DIRAC{s_2}+\frac{2}{3}\cdot\DIRAC{s_3}$, which
cannot be simulated by $\DIRAC{s_0}$.
Intuitively, states $s_0$, $s_1$, $s_2$, and $s_3$ have the same set
of enabled actions, and both $s_0$ and $s_1$ can perform either an
$\alpha$ transition evolving into
$\{r_2:\frac{1}{3},r_3:\frac{2}{3}\}$, or perform a $\beta$ transition
leading to $\DIRAC{r_1}$. For schedulers with limited power like
partial information schedulers, $s_0$ and $s_1$ cannot be
distinguished.
\end{example}

The following theorem shows that $\EWBS$ defined in
Definition~\ref{def:weak bisimulation early} is an equivalence
relation, similarly for $\LWBS$.  Moreover $\LWBS$ is strictly coarser
than $\EWBS$ which is straightforward from Definition~\ref{def:weak
bisimulation early} and \ref{def:weak bisimulation late}.
\begin{theorem}\label{thm:equivalence relation}
  \begin{inparaenum}[\scriptsize$(i)$]
  \item $\EWBS$ and $\LWBS$ are equivalence relations;
  \item $\EWBS~\subset~\LWBS$.
  \end{inparaenum}
\end{theorem}

\section{Observable Behaviour and Composition}\label{sec:observable}

In this section we consider important properties of late weak
bisimulation, namely preservation of trace distributions, and
compositionality. While these properties do \emph{not} hold if
considering all schedulers, we establish them for the subclass of
partial information distributed schedulers. Partial information
schedulers $\schedClassPartialInformation$ have been coined by De
Alfaro~\cite{de1999verification}, and distributed schedulers 
$\schedClassDistributedSchedulers$ stem from~D'Argenio and Giro
\cite{Giro2007QMC}. Both have been proposed to rule out unrealistic scheduling
decisions such as the ones discussed in Fig.~\ref{fig:motivation}. We
echo these arguments to back our claim that late weak bisimulation is
a valuable relation in the context of any realistic scheduling.  To
get started, we review desirable properties we are going to discuss.
For this, we recall the parallel operator introduced
in~\cite{eisentraut2010probabilistic}. It is an entirely
straightforward adaptation of parallel composition for $\IMC$s and for
$\PA$s.
\begin{definition}
\label{def:parallel operator}
  Let $\ML_1=\aMA[{}_1]$ and
  $\ML_2=\aMA[{}_2]$ be two $\MA$s, and $A \subseteq \ACT$ then
  $\APAR{\ML_1}{\ML_2}=\aMA[]$
  such that $\initialState = \APAR{\initialState_1}{\initialState_2}$, 
  \begin{inparaenum}[\scriptsize$(i)$]
  \item  $\stateSet=\{\APAR{s_1}{s_2}\mid
    (s_1,s_2)\in\stateSet_1\times\stateSet_2\}$, 
  \item $(\APAR{s_1}{s_2},\alpha,\APAR{\mu_1}{\mu_2})\in
    \PTRAN{}$ iff either $\alpha\in A$ and
    $s_i\PTRAN{\alpha}_i\mu_i$ for all $i\in\{1,2\}$ or
    $\alpha\notin A$, $s_i\PTRAN{\alpha}_i\mu_i$, and
    $\mu_{3-i}=\DIRAC{s_{3-i}}$ for  $i\in\{1,2\}$,
  \item \ls{changed} $(\APAR{s_1}{s_2},\lambda,\APAR{s'_1}{s'_2})\in\MTRAN{}$ iff
   either $s_i=s'_i$ and $(s_i,\lambda_i,s'_i)\in\MTRAN{}_i$ 
   with $\lambda_1+\lambda_2=\lambda$, 
   or $(s_i,\lambda, s'_i)\in\MTRAN{}_i$ and $s_{3-i}=s'_{3-i}$ for $i\in\{1,2\}$,
  \end{inparaenum}
  where $\APAR{\mu_1}{\mu_2}$ is a distribution such that
  $(\APAR{\mu_1}{\mu_2})(\APAR{s_1}{s_2})=\mu_1(s_1)\cdot\mu_2(s_2)$.
\end{definition}

We now introduce the notion of trace distribution
equivalence~\cite{Segala1995CTS} adapted to our setting\hh{better
  would be timed trace distributions actually! But not for this
  submission I suppose.}.  Let $\TRACE\in(\ACT\cup\RPLUS)^{*}$ denote
a finite trace of an $\MA$ $\ML$ consisting of an ordered sequence
of visible actions. Moreover, the cylinder $C_{\TRACE}$ induced by
$\TRACE$ is defined by: $C_{\TRACE}=\cup\{\MI{Cyl}(\Pi)\mid
\Pi\in\MF{F}_{\PATHS^{*}}\land\MI{trace}(\Pi)=\TRACE\}$ where
$\MI{trace}(\Pi) = \epsilon$ denoting an empty trace if $\ABS{\Pi}\leq 1$, and
$
\MI{trace}(\Pi) =
\begin{cases} 
 \MI{trace}(\Pi') & \Pi =
\Pi'\circ(\theta,t,s')\land\theta\in\{\tau,0\} \\ \MI{trace}(\Pi')\theta & \Pi =
\Pi'\circ(\theta,t,s')\land\theta\not\in\{\tau,0\}
\end{cases}.
$
The measurability of $C_{\TRACE}$ is straightforward from its
definition since it is a countable set of  cylinders
$\MI{Cyl}(\Pi)$.  Below we define a family of equivalences, 
parametrized by certain classes of schedulers.\ls{Use a slight different
symbol for schedulers, since the previous symbol has already been used for states.}
\begin{definition}
Let
$s_1$ and $s_2$ be two states of an $\MA$, and $\schedClass$ a set
of schedulers. Then, $s_1~\TREQUIV[\schedClass]~s_2$ iff for each scheduler $\SCH_1\in\schedClass$
there exists a scheduler $\SCH_2\in\schedClass$, such that
$\PR_{s_1}^{\SCH_1}(C_{\TRACE})=\PR_{s_2}^{\SCH_2}(C_{\TRACE})$ for each
finite trace $\TRACE$  and
vice versa. If $\schedClass$ is the set of all schedulers, we
simply write $\TREQUIV$.
\end{definition} 
 Below follow examples (and counterexamples) of trace distribution equivalent states:

\begin{example}\label{ex:trace} Let $s_0$ and $s_0'$ be two states in
Fig.~\ref{fig:motivation}, then we have $s_0~\TREQUIV~s_0'$, since the
only trace distribution of $s_0$ and $s_0'$ is
$\{ih:\frac{1}{2},it:\frac{1}{2}\}$.  In contrast, 
$s_0$ and $s_1$ in Fig.~\ref{fig:pa} are not trace distribution
equivalence. Since there are two possible trace distributions for
$s_0$: $\{\beta:1\}$ and $\{\alpha:1\}$, but for $s_1$ there are four
trace distributions: $\{\alpha:1\}$, $\{\beta:1\}$,
$\{\alpha:\frac{1}{3},\beta:\frac{2}{3}\}$, and
$\{\beta:\frac{1}{3},\alpha:\frac{2}{3}\}$.
\end{example}

\subsection{Realistic  Schedulers}

We are now refining the very liberal Definition~\ref{def:scheduler}
where the set of all schedulers was introduced. As discussed, this
class can be considered too powerful, since it  includes unrealistic
schedules such as the one scheduling the coloured execution of $\APAR{s'_0}{r_0}$
depicted in Fig.~\ref{fig:execution}.

In the following we define two prominent sub-classes of schedulers,
where only limited information is at hand for scheduling.  We need to
first introduce some notations. Let $\EACT:\stateSet\mapsto
2^{\ACT\cup\RPLUS}$ such that
$\EACT(s)=\{\theta\in(\ACT\cup\RPLUS)\mid
\exists\mu.s\WTRAN{\theta}\mu\},$ that is, the function $\EACT$
returns the set of visible actions that a state is able to perform,
possibly after some internal transitions.  We generalize this
function to paths as follows: $\EACT(\pi) = \EACT(s)$ if $\pi=s$, and $\EACT(\pi)=$
{\small
\begin{numcases}{}
  \EACT(\pi')&$\pi=\pi'\circ(\alphaup,t,s)\land\alphaup\in\{\tau,0\}\land\EACT(\LAST[\pi'])=\EACT(s)$\label{case:eact 1}\\
  \EACT(\pi'),\alphaup,t,\EACT(s) & $\pi=\pi'\circ(\alphaup,t,s)\land(\alphaup\not\in\{\tau,0\}\lor\EACT(\LAST[\pi'])\neq\EACT(s))$
\end{numcases}}
where Case~(\ref{case:eact 1}) takes care of a special situation such that
internal actions do not change enabled actions. In this case $\EACT$
will not see the difference.  Intuitively, $\EACT(\pi)$ abstracts
concrete states on $\pi$ to their corresponding enabled
actions. Whenever an invisible action does not change the enabled
actions, this will simply be omitted.
In other words, $\EACT(s)$ can be seen as the interface of $s$,
which is observable by other components. Other components
can observe the execution of $s$, as long as either it performs
a visible action ($\theta\not\in\{\tau,0\}$), or its interface
has been changed ($\EACT(\LAST[\pi']\neq\EACT(s)$).
We are now ready to define the \emph{partial information schedulers}~\cite{de1999verification} as follows:
\begin{definition}
\label{def:pis}
  A scheduler $\SCH$ is a partial information scheduler of $s$ if for any $\pi_1,\pi_2\in\PATHS^{*}(s)$,
  $\EACT(\pi_1)=\EACT(\pi_2)$ implies:
  \begin{inparaenum}[\scriptsize$(i)$]
  \item either $\SCH(\pi_1)=(\tau,\mu)$ or $\SCH(\pi_2)=(\tau,\mu)$
    for some $\mu$,
  \item or $\SCH(\pi_1)=(\theta,\mu)$ and $\SCH(\pi_2)=(\theta,\nu)$ for some $\mu,\nu$ such that $\theta\neq\tau$.
  \end{inparaenum}
\end{definition}
We denote the set of all partial information schedulers by
$\schedClassPartialInformation$.  Intuitively a partial information
scheduler can only distinguish states via different enabled visible
actions. It therefore excludes the possibility to schedule differently
only because of different state identities. This fits very well to a
behaviour-oriented rather than state-oriented view, as it is typical
for process calculi. 
Consequently, for two different paths $\pi_1$ and $\pi_2$ with
 $\EACT(\pi_1)=\EACT(\pi_2)$, a partial information scheduler
either chooses a transition labelled with $\tau$ action for $\pi_i$
($i=1,2$), or it chooses transitions labelled with the same visible
actions for both $\pi_1$ and $\pi_2$. Partial information schedulers
do not impose any restriction on the execution of $\tau$ transitions,
instead they can be performed spontaneously.

In order to exclude unrealistic schedulers when composing parallel
systems, another important sub-class of schedulers called
\emph{distributed schedulers} has been
introduced~\cite{Giro2007QMC}. The idea of distributed schedulers is
to assume that a component running in parallel to other components
needs to make its local scheduling decisions in isolation, and thus
can use only that information about other components that has been
communicated to them beforehand. For instance the guesser in
Fig.~\ref{fig:motivation} cannot base its local scheduling decision
on the tossing outcome at the moment when his guess is to be scheduled.

To formalise this locality idea, we first need to define the
projection of a path to the path of its components. Let
$s=\ \parallel_A\{s_i\mid 1\le i\le n\}$ be a state which is composed
by $n>1$ processes in parallel such that all the processes synchronize
on actions in $A$.  Let $\pi$ be a path starting from $s$, then the
$i$-projection of $\pi$ denoted by $\PROJ{\pi}{i}$ is defined as
follows: $\PROJ{\pi}{i}=\PROJ{s}{i}$ if $\pi=s$, otherwise
$$
\PROJ{\pi}{i} =
\begin{cases}
  \PROJ{\pi'}{i}\circ(\theta,t,\PROJ{s'}{i}) & \pi=\pi'\circ(\theta,t,s')\land(\theta\in A\lor(\theta\not\in A\land\PROJ{\LAST[\pi']}{i}\TRAN{\theta}\PROJ{s'}{i}))\\
  \PROJ{\pi'}{i} & \pi=\pi'\circ(\theta,t,s')\land\theta\not\in A\land(\exists j\neq i.\PROJ{\LAST[\pi']}{j}\TRAN{\theta}\PROJ{s'}{j})\\
\end{cases}
$$
where $\PROJ{s}{i}=s_i$ with $1\le i\le n$. Intuitively, given a path
$\pi$ of a state $s$, the $i$-projection of $\pi$ is the path that
only keeps track of the execution of the $i$-th component of $s$
during its execution. Below defines the
distributed schedulers in an inductive way.

\begin{definition}
\label{def:distributed scheduler}
  A scheduler $\SCH$ is a distributed scheduler of
  $s=\parallel_A\{s_i\mid 1\le i\le n\}$ iff for any $\pi,\pi'\in\PATHS^{*}(s)$, 
  $\PROJ{\pi}{i}=\PROJ{\pi'}{i}$ for
  each $1\le i\le n$ implies $\SCH(\pi)=\SCH(\pi')$. 
\end{definition}
We denote the set of all distributed schedulers by
$\schedClassDistributedSchedulers$. In case $n=1$,
  distributed schedulers degenerate to ordinary schedulers defined in
  Definition~\ref{def:scheduler}.  According to
Definition~\ref{def:distributed scheduler}, a scheduler $\SCH$ is
distributed, if $\SCH$ cannot distinguish different paths starting
from $s$, provided the projections of these paths to each of its
parallel component coincide. Note that the scheduler inducing the coloured
execution in Fig.~\ref{fig:execution} is not distributed, since the
decision of $r_0$ depends on the execution history of $s'_0$, i.e. at
state $s_5$, $r_0$ will choose the left transition, and it will choose
the right transition while at state $s_6$. By restricting to the set
of distributed schedulers, we can avoid the unrealistic execution of
$\APAR{s'_0}{r_0}$ depicted in Fig.~\ref{fig:execution}.

\subsection{Properties of Late Weak Bisimilarity}
\label{sec:properties}
In this section we show properties of late weak bisimilarity
under realistic schedulers. We first introduce some notations:
Let $\distHistoryU$ and $\distHistoryV$ denote distributions over set 
$\{(\pi,s)\mid\pi\in\PATHS^{*}\land s\in\stateSet\land\LAST=s\}$,
moreover $\PROJ{\distHistoryU}{2}=\mu$ denotes the projection of
$\distHistoryU$ to its corresponding distribution of states i.e.
$\mu(s)=\sum\{\distHistoryU((\pi,s))\mid\exists\pi.(\pi,s)\in\SUPP(\distHistoryU)\}$.
Given a scheduler $\SCH$, a transition from $\distHistoryU$ to $\distHistoryV$
with label $\theta$ is induced by $\SCH$, 
written as $\distHistoryU\scheTran{\theta}{\SCH}\distHistoryV$, 
iff $\distHistoryV((\pi\circ(\theta,t,s'),s'))=\distHistoryU((\pi,s))\cdot\nu_{(\pi,s)}(s')$
where $\nu_{(\pi,s)}=\sum_{\nu\in\DIST(\stateSet)}\SCH(\pi,\theta,\nu)\cdot\nu$
for each $(\pi,s)\in\SUPP(\distHistoryU)$. 
Namely, $\nu_{(\pi,s)}$ is the resulting distribution of $s$ under
scheduler $\SCH$ given the history information $\pi$. 
For each $s'\in\SUPP(\nu_{(\pi,s)})$, the probability of $s'$ in 
$\distHistoryV$ is weighted by $\distHistoryU((\pi,s))$,
moreover we need to update the history information $\pi$ to
$\pi\circ(\theta,t,s')$.
Correspondingly, a transition from $\mu$ to $\nu$ with label $\theta$ is induced by a scheduler $\SCH$,
written as $\mu\scheTran{\theta}{\SCH}\nu$, iff 
$\distHistoryU\scheTran{\theta}{\SCH}\distHistoryV$ such that 
$\PROJ{\distHistoryV}{2}=\nu$, where $\distHistoryU((s,s))=\mu(s)$ 
for each $s\in\SUPP(\mu)$. Intuitively, given a distribution $\mu$, for each
$s\in\SUPP(\mu)$ we use $s$ as the history information for $\SCH$ to guide
the execution, since it is the only priori information we have known so far.
Similarly, we can define weak transitions of $\mu$ induced
by a given scheduler.
Based on the notations introduced above, we can modify Definition~\ref{def:weak bisimulation late}
with schedulers being considered explicitly.

\begin{definition}
\label{def:weak bisimulation late schedulers}
Let $\SCH_1,\SCH_2,\SCH\in\schedClass$
for a given set of schedulers $\schedClass$. A relation
$\MC{R}\subseteq\DIST(\stateSet)\times\DIST(\stateSet)$ is a late weak
bisimulation over $\ML$ with respect to $\schedClass$ 
iff $\mu~\MC{R}~\nu$ implies:
  \begin{inparaenum}[\scriptsize$(i)$]
  \item whenever $\mu\scheTran{\theta}{\SCH_1}\mu'$, 
    there exists $\nu\scheWTran{\theta}{\SCH_2}\nu'$ such that 
    $\mu'~\MC{R}~\nu'$;
  \item if not $\TC{\mu}$, then there exists 
  $\mu=\sum_{0\leq i\leq n}p_i\cdot\mu_i$  and 
  $\nu\scheWTran{\tau}{\SCH}\sum_{0\leq i\leq n}p_i\cdot\nu_i$ 
  such that $\TC{\mu_i}$ and $\mu_i~\MC{R}~\nu_i$ 
  for each $0\leq i\leq n$ where $\sum_{0\leq i\leq n}p_i=1$;
\item symmetrically for $\nu$.
\end{inparaenum}
 The meanings of $\mu~\LWBS_{\schedClass}~\nu$ and $s~\LWBS_{\schedClass}~r$
are the same as in Definition~\ref{def:weak bisimulation late}.
\end{definition}

Definition~\ref{def:weak bisimulation late schedulers} is almost the same as 
Definition~\ref{def:weak bisimulation late} except that we require every transition is induced by a scheduler in $\schedClass$. As we shall
prove later, these two definitions are actually equivalent.

As mentioned before, late weak bisimulation has a flavour similar
to partial information schedulers in the sense that, due to the
transition consistency requirement, there is no difference
between states in the support of a distribution if the same set of
actions is enabled. Indeed, late weak bisimulation and partial
information schedulers are closely related. The following theorem
states that partial information schedulers are enough to discriminate
late weak bisimilarity with respect to arbitrary schedulers, and that
if restricting to partial information schedulers, late
weak bisimulation implies trace distribution equivalence.

\begin{theorem}\label{thm:implication}
  For any states $s_1$ and $s_2$, 
 $
s_1~\LWBS~s_2 \Longleftrightarrow s_1~\LWBS_{\schedClassPartialInformation}~s_2 \Longrightarrow s_1~\TREQUIV[\schedClassPartialInformation]~s_2.
$
\end{theorem}

Theorem~\ref{thm:implication} does not hold if we consider general
schedulers:
\begin{example} Let $s_0$ and $s_1$ be two states in
Fig.~\ref{fig:pa}, and in Example~\ref{ex:pa} we have shown that
$s_0~\LWBS~s_1$, while in Example~\ref{ex:trace} we have shown that
$s_0~\not\TREQUIV~s_1$.  But we also notice that the schedulers giving
rise to the trace distributions
$\{\alpha:\frac{1}{3},\beta:\frac{2}{3}\}$ and
$\{\beta:\frac{1}{3},\alpha:\frac{2}{3}\}$ are not partial information
schedulers, since at states $s_2$ and $s _3$ with the same enabled
visible actions, the schedulers can choose transitions with different
labels. By restricting to partial information schedulers we 
exclude these two distributions and can indeed show that $s_0~\TREQUIV[\schedClassPartialInformation]~s_1$.
\end{example}

It is worthwhile to recall that $\MA$ have a continuous time
semantics, thus trace distribution equivalence implicitly relates
the timed probabilistic behaviour of an $\MA$, basically because traces 
are composed of external actions as well as rates, hence rates are equated by trace equivalence\hh{can we say this a bit clearer?}\ls{pls check}. This implies that
for instance timed reachability probabilities are preserved. So, if we
let $\PR^{\SCH}_s(\diamondsuit^{\le t}G)=
\PR_{s}^{\SCH}(\{\pi\in\PATHS^{\omega}\mid\exists n\geq 0.(\pi[n]\in
G\land\TTIME(\pi[0..n])\leq t)\})$ denote the probability of reaching
states in $G$ from $s$ in no more than $t$ time units, under scheduler
$\SCH$, we can establish that $s_1~\LWBS~s_2$ implies for arbitrary
scheduler $\SCH_1\in\schedClassPartialInformation$, there exists
$\SCH_2\in\schedClassPartialInformation$ such that
$\PR^{\SCH_1}_{s_1}(\diamondsuit^{\le t}G) =
\PR^{\SCH_2}_{s_2}(\diamondsuit^{\le t}G)$.  

If looking at the effect of parallel composition, we need to restrict
to distributed schedulers to establish compositionality, as indicated
by the following theorem:
\begin{theorem}\label{thm:distributed composition}
  For two states $s_1$ and $s_2$ of an \MA{}, 
  \begin{enumerate}
    \item $s_1~\LWBS~s_2 \Longleftrightarrow s_1~\LWBS_{\schedClassDistributedSchedulers}~s_2$, 
      provided $s_1$ and $s_2$ are sequential i.e.
   contain no parallel operators;
  \item $s_1~\LWBS_{\schedClassDistributedSchedulers}~s_2 \Longrightarrow
  \APAR{s_1}{s_3}~\LWBS_{\schedClassDistributedSchedulers}~\APAR{s_2}{s_3}$ for any $s_3$. 
  \end{enumerate}
\end{theorem}

In Clause~1 of Theorem~\ref{thm:distributed composition}, we require that both $s_1$ and $s_2$ 
contain no parallel operators, otherwise the implication does not hold. 
Moreover when general schedulers are considered,
Clause~2 of Theorem~\ref{thm:distributed composition} will not hold either.
This is demonstrated by the following two examples:

\begin{example}
  Let $\APAR{s'_0}{r_0}$ be a state as in Example~\ref{ex:motivation}, whose execution is
  depicted in Fig.~\ref{fig:execution}~(b). Additionally, let $r$ be a sequential
  state whose execution is same as $\APAR{s'_0}{r_0}$, such sequential state always
  exists (simply introducing a state for each node in Fig.~\ref{fig:execution} (b)).
  By construction, we have $\APAR{s'_0}{r_0}~\LWBS~r$. 
  However, if restricted to schedulers in $\schedClassDistributedSchedulers$, 
  $\APAR{s'_0}{r_0}~\LWBS_{\schedClassDistributedSchedulers}~r$ does not hold. Since
  the scheduler inducing the coloured execution of $\APAR{s'_0}{r_0}$ in Fig.~\ref{fig:execution} (b)
  is not distributed, while the scheduler inducing the corresponding execution of $r$
  is distributed. Essentially, every possible scheduler of $r$ is distributed because $r$
  is sequential.
\end{example}

\begin{example}\label{ex:distributed} Let $s_0$, $s'_0$, and $r_0$ be
the states in Fig.~\ref{fig:motivation}. We have shown
in Example~\ref{ex:ex1} that $s_0~\LWBS~s'_0$, but we have
$\APAR{s_0}{r_0}~\not\LWBS~\APAR{s'_0}{r_0}$ if general schedulers are
considered. Since the coloured execution of $\APAR{s'_0}{r_0}$ depicted in
Fig.~\ref{fig:execution} cannot be simulated by $\APAR{s_0}{r_0}$ no
matter how we schedule the transitions of $\APAR{s_0}{r_0}$. For
instance the probability for $\APAR{s'_0}{r_0}$ reaching states
$\APAR{s_3}{r_5}$ and $\APAR{s_4}{r_6}$ is equal to 1, while the
probability for $\APAR{s_0}{r_0}$ reaching these two states is at most 0.5.

However, when restricting to distributed schedulers, we can show that
both \mbox{$\APAR{s_0}{r_0}$} and $\APAR{s'_0}{r_0}$ can reach states
$\APAR{s_3}{r_5}$ and $\APAR{s_4}{r_6}$ with probability 0.5 at most,
since the scheduler of $\APAR{s'_0}{r_0}$, which induces the coloured
execution in Fig.~\ref{fig:execution} is not distributed. The reason
is that at states $\APAR{s_5}{r_0}$ and $\APAR{s_6}{r_0}$, $r_0$ makes different 
decision by looking at the future transitions of $s_5$ and $s_6$, 
which should not happen in a distributed scheduler.
\end{example}

When restricting to the set of
schedulers in $\schedClassPartialInformation\cap\schedClassDistributedSchedulers$, 
late weak bisimulation is compositional and implies trace distribution equivalence.
Actually, we can show  that with respect to schedulers in
$\schedClassPartialInformation\cap\schedClassDistributedSchedulers$,
late weak bisimulation is the coarsest congruence preserving trace
distribution equivalence,
which in turn can be seen 
as the symmetric version of trace distribution precongruence defined 
in~\cite{Lynch2007OBS}.

\begin{theorem}\label{thm:trace congruence}
  Let $\schedClass=\schedClassPartialInformation\cap\schedClassDistributedSchedulers$, 
  then $s_1~\LWBS_{\schedClass}~s_2$ iff $s_1~\traceCongruence~s_2$ for any $s_1$ and $s_2$,
  where $s_1~\traceCongruence~s_2$ iff $s_1~\TREQUIV[\schedClass]~s_2$ and 
  $\APAR{s_1}{s_3}~\TREQUIV[\schedClass]~\APAR{s_2}{s_3}$ for any 
  $s_1,s_2,s_3$, and $A$.
\end{theorem}

\section{Conclusion and Future Work}
\label{sec:conclusion} 
In this paper, we have presented a novel and very coarse weak
bisimilarity called late weak bisimilarity for $\MA$s. Late weak
bisimilarity has interesting properties under two well-known
subclasses of schedulers: It implies trace distribution equivalence
under partial information schedulers, while it is compositional 
under distributed
schedulers. Working in the intersection of both scheduler classes thus
ensures a restricted form of compositionality, where the restriction
excludes undesired or unrealistically powerful schedulers.
As future work we intend to study reduction barbed
congruences~\cite{dengsemantics} under subclasses of schedulers, in order
to pinpoint the characteristics of late weak bisimilarity. 
The logical characterization of $\LWBS$ would be also interesting.
Moreover, we are working on an efficient decision algorithm for $\LWBS$. 
We expect that
the decision algorithm for $\LWBS$ is simpler than the algorithm for
$\EWBS$, since we do not allow arbitrary splitting, thus 
it is enough to consider all reachable transition consistent distributions, 
which are finitely many. However, this is not the case for $\EWBS$. To the best
of our knowledge, the most efficient algorithm so far to decide $\EWBS$
is exponential, see~\cite{SchusterSMA2013,QEST2013}.

\newpage
\appendix

\section{Proofs}
\subsection{Proof of Theorem~\ref{thm:equivalence relation}}
Before proving Theorem~\ref{thm:equivalence relation}, we shall introduce
two lemmas.
In Definition~\ref{def:weak bisimulation early} and
\ref{def:weak bisimulation late} we have used strong transitions on
the left side of Clause~$(\ref{item:late 1})$. As in the standard setting for transition
systems, in the lemma below we show that weak bisimulation does not
change if we replace the strong transition by weak transition. This
simple replacement is useful for proving the transitivity.
\begin{lemma}\label{lem:weak bisimulation same} Let
$\ML=\aMA{}$ be an $\MA$.  A relation
$\MC{R}\subseteq\DIST(\stateSet)\times\DIST(\stateSet)$ is an early weak
bisimulation iff $\mu~\MC{R}~\nu$ implies that
  \begin{enumerate}
  \item whenever $\mu\WTRANP{\theta}\mu'$, there exists a
$\nu\WTRANP{\theta}\nu'$ such that $\mu'~\MC{R}~\nu'$,
  \item whenever $\mu=\sum_{0\leq i\leq n}p_i\cdot\mu_i$, there exists
$\nu\WTRANP{\tau}\sum_{0\leq i\leq n}p_i\cdot\nu_i$ such that
$\mu_i~\MC{R}~\nu_i$ for each $0\leq i\leq n$ where $\sum_{0\leq i\leq
n}p_i=1$,
  \item symmetrically for $\nu$.
  \end{enumerate} Similar results hold for late weak bisimulation by
adjusting Clause 2 accordingly.
\end{lemma}
\begin{proof} Note that $\mu\WTRANP{\theta}\mu'$ iff for each
$s\in\SUPP(\mu)$, there exists $s\WTRANP{\theta}\mu_s$ such that
$\mu'=\sum_{s\in\SUPP(\mu)}\mu(s)\cdot\mu_s$.  Define
$s\WTRAN{\theta,n}\mu_s$ inductively as follows:
 \begin{enumerate}
  \item $\mu_s=\DIRAC{s}$ if $n=0$ and $\theta=\tau$,
  \item If $n>0$, then either there exists $s\TRANP{\tau}\nu$ such
that $s'\WTRANP{\theta,(n-1)}\mu_{s'}$, or there exists
$s\TRAN{\theta}\nu$ such that $s'\WTRAN{\tau,(n-1)}\mu_{s'}$ for
each $s'\in\SUPP(\nu)$, where
$\mu_s=\sum_{s'\in\SUPP(\nu)}\nu(s')\cdot\mu_{s'}$.
  \end{enumerate} In other words, $s\WTRAN{\theta,n}\mu_s$ means that
$\mu_s$ can be reached in $n$ steps from $s$. Similarly, we can define
$\mu\WTRAN{\theta,n}\mu'$.

  We first prove that whenever $\mu\WTRANP{\theta,n}\mu'$, there
exists $\mu\WTRANP{\theta}\nu'$ such that $\mu'~\MC{R}~\nu'$ which can
be done by induction on $n$:
  \begin{enumerate}
  \item $n=0$. Trivial, since it must be the case that $\theta=\tau$
and $\mu'=\mu$.
  \item $n>0$. Then there exists either
$\mu\TRANP{\theta}\mu_1\WTRANP{\tau,(n-1)}\mu'$, or
$\mu\TRANP{\tau}\mu_1\WTRANP{\theta,(n-1)}\mu'$.  We only show the
proof of the first case, since the other one is similar.  By
Definition~\ref{def:weak bisimulation early}, there exists
$\nu\WTRANP{\theta}\nu_1$ such that $\mu_1~\MC{R}~\nu_1$.  The
following proof is by induction hypothesis showing that there exists
$\nu_1\WTRANP{\tau}\nu'$ such that $\mu'~\MC{R}~\nu'$.  Consequently,
there exists $\nu\WTRANP{\theta}\nu'$ such that $\mu'~\MC{R}~\nu'$ as
desired.
  \end{enumerate}

Next we show that $\mu\WTRANP{\tau}\mu'$ iff there exists
$\{\mu\WTRANP{\tau,n}\mu'_n\}_{n\ge 0}$ such that $\mu' =
\lim_{n\rightarrow\infty}\mu'_n$. It suffices to show that
for any $n\ge 0$ and $s$, whenever 
\begin{align*}
 \DIRAC{s} &= \mu_0^{\to} + \mu_0^{\times}, \\
\mu_0^{\to} &\TRAN{\tau} \mu_1^{\to} +\mu_1^{\times},\\
\mu_1^{\to} &\TRAN{\tau} \mu_2^{\to} + \mu_2^{\times},\\
&\vdots \\
\mu_{n-1}^{\to} &\TRAN{\tau} \mu_n^{\times},
\end{align*}
there exists $\DIRAC{s}\WTRANP{\tau,n}\mu_s\equiv(\sum_{0\le i \le n}\mu_i)$. 
This can be proved by induction on $n$:
\begin{enumerate}
\item $n=0$. Trivial.
\item $n>0$. By induction hypothesis, $\DIRAC{s}\WTRANP{\tau,n-1}\mu'_s$,
where $\mu'_s\equiv(\mu^\to_{n-1}+\sum_{0\le i<n}\mu^{\times}_i)$.
Since $\mu^\to_{n-1}\TRAN{\tau}\mu^{\times}_n$, there
exists $\DIRAC{s}\WTRANP{\tau,n}\mu_s\equiv(\sum_{0\le i\le n}\mu^{\times}_i)$.
\end{enumerate}

We have proved that $\mu\WTRANP{\tau}\mu'$ iff there exists
$\{\mu\WTRANP{\tau,n}\mu'_n\}_{n\ge 0}$ such that $\mu' =
\lim_{n\rightarrow\infty}\mu'_n$. Therefore we can conclude that
whenever $\mu\WTRANP{\tau}\mu'$, there exists
$\nu\WTRANP{\tau}\nu'$ such that $\mu'~\MC{R}~\nu'$. 
In case $\mu\WTRANP{\theta}\mu'$ with $\theta\neq\tau$,
we have $\mu\WTRANP{\tau}\mu'_1\TRANP{\theta}\mu'_2\WTRANP{\tau}\mu'$.
As shown above, there exists $\nu\WTRANP{\tau}\nu'_1$ such that $\mu'_1~\MC{R}~\nu'_1$,
which indicates that there exists $\nu'_1\WTRANP{\theta}\nu'_2$ such that $\mu'_2~\MC{R}~\nu'_2$
by Definition~\ref{def:weak bisimulation early}, which indicates that 
there exists $\nu'_2\WTRANP{\tau}\nu'$ such that $\mu'~\MC{R}~\nu'$.  This completes
the proof.  
\end{proof}

In order to prove that $\LWBS$ is an equivalence relation, we shall
introduce the following lemma saying that if $\mu~\LWBS~\nu$, then
$\mu$ and $\nu$ must be transition consistent or not at the same time.
\begin{lemma}\label{lem:consistent preserved} For all late weak
bisimulation $\MC{R}$, $\mu~\MC{R}~\nu$ implies $\TC{\mu}$ iff
$\TC{\nu}$.
\end{lemma}
\begin{proof} We prove by contradiction and assume $\mu~\MC{R}~\nu$
and $\TC{\mu}$ for some late weak bisimulation $\MC{R}$, but not
$\TC{\nu}$.  Since $\mu~\MC{R}~\nu$, then $\mu\WTRAN{\theta}$ implies
$\nu\WTRAN{\theta}$ and vice versa for any $\theta$, therefore we have
$\EACT(\mu)=\EACT(\nu)$, where
$\EACT(\mu)=\{\theta\mid\exists\mu'.\mu\WTRAN{\theta}\mu'\}$,
similarly for $\EACT(\nu)$. Since $\nu$ is not transition consistent,
there exists $s\in\SUPP(\nu)$, such that $s\WTRAN{\theta}$ where
$\theta\not\in \EACT(\nu)$.  Therefore there exists $\nu=\sum_{i\in
I}p_i\cdot\nu_i$ such that $\TC{\nu_i}$ for each $i\in I$ and there
exists $j\in I$ such that $\nu_j\WTRAN{\theta}$, where $I$ is a finite
set of indexes.  Since $\TC{\mu}$ and $\theta\not\in \EACT(\mu)$,
there does not exist $\mu\WTRAN{\tau}\sum_{i\in I}p_i\cdot\mu_i$ such
that $\mu_i\WTRAN{\theta}$, and thus $\mu_i~\MC{R}~\nu_i$, which
contradicts the assumption that $\mu~\LWBS~\nu$.  
\end{proof}

Below follows the proof of Theorem~\ref{thm:equivalence relation}:
\begin{proof} The proofs for early weak bisimulation is
straightforward from Lemma~\ref{lem:weak bisimulation same} and
omitted here.  We prove that $\LWBS$ is an equivalence relation. The
only non-trivial case is transitivity, we need to prove that
$\mu~\LWBS~\nu$ and $\nu~\LWBS~\gamma$ implies $\mu~\LWBS~\gamma$ for
any $\mu,\nu$, and $\gamma$.  According to Definition~\ref{def:weak
bisimulation late}, there exists late weak bisimulations $\MC{R}_1$
and $\MC{R}_2$ such that $\mu~\MC{R}_1~\nu$ and $\nu~\MC{R}_2~\gamma$.
Let
$\MC{R}=\MC{R}_1\circ\MC{R}_2=\{(\mu,\gamma)\mid\exists\nu.(\mu~\MC{R}_1~\nu\land
\nu~\MC{R}_2~\gamma)\}$, it is then enough to prove that $\MC{R}$ is
also a late weak bisimulation.

  Let $\mu~\MC{R}~\gamma$ such that $\mu~\MC{R}_1~\nu$ and
$\nu~\MC{R}_2~\gamma$ for some $\nu$.  First we prove that whenever
$\mu\WTRANP{\theta}\mu'$, there exists $\gamma\WTRANP{\theta}\gamma'$
such that $\mu'~\MC{R}~\gamma'$. Due to Lemma~\ref{lem:consistent
preserved}, the proof is straightforward and omitted here.

  Secondly, we need to show that if not $\TC{\mu}$, then there exists
$\mu=\sum_{i\in I}p_i\cdot\mu_i$ and $\gamma\WTRANP{\tau}\sum_{i\in
I}p_i\cdot\gamma_i$ such that $\mu_i~\MC{R}~\gamma_i$ for each $i\in
I$, where $\sum_{i\in I}p_i=1$.  Since $\mu~\LWBS~\nu$, there exists
$\nu\WTRANP{\tau}\sum_{i\in I}p_i\cdot\nu_i$ such that $\TC{\mu_i}$
and $\mu_i~\MC{R}_1~\nu_i$ for each $i\in I$. By
Lemma~\ref{lem:consistent preserved}, $\TC{\nu_i}$ for each $i\in
I$. We distinguish the following two cases:
  \begin{enumerate}
  \item $\nu=\sum_{i\in I}p_i\cdot\nu_i$.\\ According to
Lemma~\ref{lem:consistent preserved}, $\nu$ is not transition
consistent, and moreover we have $\TC{\nu_i}$ for each $i\in I$.
Since $\nu~\MC{R}_2~\gamma$, there exists
$\gamma\WTRANP{\tau}\sum_{i\in I}\gamma_i$ such that
$\nu_i~\MC{R}_2~\gamma_i$, thus we have $\mu_i~\MC{R}~\gamma_i$ by the
definition of $\MC{R}$ for each $i\in I$.
  \item $\nu\WTRANP{\tau}\nu'=\sum_{i\in I}p_i\cdot\nu_i$. \\ Since
$\nu~\MC{R}_2~\gamma$, there exists $\gamma\WTRANP{\tau}\gamma'$ such
that $\nu'~\MC{R}_2~\gamma'$ according to the first clause of
Definition~\ref{def:weak bisimulation late}.  Since $\mu$ is not
transition consistent, so there exists $i,j\in I$ such that $i\ne j$
and $\EACT(\mu_i)\ne\EACT(\mu_j)$, which indicates that
$\EACT(\nu_i)\ne\EACT(\nu_j)$, therefore $\nu'$ is not transition
consistent. As a result there exists $\gamma'\WTRANP{\tau}\sum_{i\in
I}p_i\cdot\gamma_i$ i.e.  $\gamma'\WTRANP{\tau}\sum_{i\in
I}p_i\cdot\gamma_i$ such that $\nu_i~\MC{R}_2~\gamma_i$, thus
$\mu_i~\MC{R}~\gamma_i$ for each $i\in I$.
  \end{enumerate}

  For Clause 2, it is easy to see that the second condition of
Definition~\ref{def:weak bisimulation early} implies the second
condition of Definition~\ref{def:weak bisimulation late}, but not vice
versa. Example~\ref{ex:ex1} shows that the inclusion is strict.  
\end{proof}

\subsection{Proof of Theorem~\ref{thm:implication}}
\begin{proof}
  \begin{enumerate}
  \item $s_1~\LWBS~s_2 \Longleftrightarrow s_1~\LWBS_{\schedClassPartialInformation}~s_2$:\\
    This equivalence is straightforward from Definition~\ref{def:weak bisimulation late},
    since we always group states with the same enable visible actions together and
    let them either perform transitions with the same visible action at the same time, or
    an internal transition spontaneously, which never breaks the conditions of 
    partial information schedulers. In other words, all transitions we consider in Definition~\ref{def:weak bisimulation late}
    are induced by some schedulers in $\schedClassPartialInformation$.
  \item $s_1~\LWBS~s_2 \Longrightarrow s_1~\TREQUIV[\schedClassPartialInformation]~s_2$:\\
Let $\mu$ and $\nu$ be two distributions such that
$\SUPP(\mu)=\{s_i\}_{i\in I}$ and $\SUPP(\nu)=\{r_i\}_{i\in J}$ where
$I$ and $J$ are two finite sets of indexes.  Let $\{\pi_i\}_{i\in I}$
and $\{\pi'_i\}_{i\in J}$ be two sets of finite paths such that
$\LAST[\pi_i]=s_i$ and $\LAST[\pi_j]=r_j$ for each $i\in I$ and $j\in
J$.  We prove a more general result: $\mu~\LWBS~\nu$ implies for each
partial information scheduler $\SCH_1$, there exists a partial
information scheduler $\SCH_2$ such that
  $$\PR_{\SCH_1}^{\mu}(C_{\TRACE},\{\pi_i\}_{i\in I})=\PR_{\SCH_2}^{\nu}(C_{\TRACE},\{\pi'_i\}_{i\in J})$$ 
  for each finite trace $\TRACE$, provided the following conditions hold:
  \begin{enumerate}
  \item $\EACT(s_i)=\EACT(s_j)$ implies $\EACT(\pi_i)=\EACT(\pi_j)$
for each $i,j\in I$,
  \item $\EACT(r_i)=\EACT(r_j)$ implies $\EACT(\pi'_i)=\EACT(\pi'_j)$
for each $i,j\in J$,
  \item $\EACT(s_i)=\EACT(r_j)$ implies $\EACT(\pi_i)=\EACT(\pi'_j)$
for each $i\in I$ and $j\in J$,
  \end{enumerate} where $\PR_{\SCH_1}^{\mu}(C_{\TRACE},\{\pi_i\}_{i\in
I})$ is the probability of $C_{\TRACE}$ starting from $\mu$ given
execution history $\pi_i$ for each $s_i\in\SUPP(\mu)$ and scheduler
$\SCH_1$.

  Since $\EACT(\pi_i)=\EACT(\pi_j)$ if $\EACT(s_i)=\EACT(s_j)$ for any
$i,j\in I$, if $\TC{\mu}$ and $\SCH_1$ is a partial information
scheduler,
  $$\PR_{\SCH_1}^{\mu}(C_{\TRACE},\{\pi_i\}_{i\in I})= \PR_{\SCH_1}^{\mu}(C_{\TRACE},\EACT(\pi_i))$$
  for any $i\in I$.  We then define
$p=\PR_{\SCH_1}^{\mu}(C_{\TRACE},\EACT(\pi_i),n)$ as follows where
$n\ge 0$ and $\TC{\mu}$:
  \begin{enumerate}
  \item If $|\TRACE|>0$ and $n=0$, $p=0$,
  \item else if $|\TRACE|=0$, $p=1$,
  \item else if $\mu\TRANPsingle{\tau}\sum_{k\in K}p_k\cdot\mu_k$ such that
$\TC{\mu_k}$ for each $k\in K$, then
    $$p  = \sum_{k\in K}p_k\cdot\PR_{\SCH_1}^{\mu_k}(C_{\TRACE},\EACT(\pi_i\circ(\tau,0,s_k)),n-1)$$
    for any $s_k\in\SUPP(\mu_k)$,
  \item else if $\TRACE=\alpha\TRACE'$ and
$\mu\TRANP{\alpha}\sum_{k\in K}p_k\cdot\mu_k$ such that $\TC{\mu_k}$
for each $k\in K$, then
    $$p  = \sum_{k\in K}p_k\cdot\PR_{\SCH_1}^{\mu_k}(C_{\TRACE'},\EACT(\pi_i\circ(\alpha,0,s_k)),n-1)$$
    for any $s_k\in\SUPP(\mu_k)$,
  \item else if $\TRACE=\lambda\TRACE'$ and
$\mu\TRANPsingle{\lambda}\sum_{k\in K}p_k\cdot\mu_k$ such that $\TC{\mu_k}$
for each $k\in K$, then
    $$p  = \sum_{k\in K}p_k\cdot\int_0^{\infty}\lambda e^{\lambda x}\cdot\PR_{\SCH_1}^{\mu_k}(C_{\TRACE'},\EACT(\pi_i\circ(\lambda,x,s_k)),n-1)dx$$
    for any $s_k\in\SUPP(\mu_k)$,
  \item otherwise $p=0$.
  \end{enumerate} If $\neg\TC{\mu}$ and $\mu=\sum_{k\in K}\mu_k$ such
that $\TC{\mu_k}$ for each $k\in K$,
then $$\PR_{\SCH_1}^{\mu}(C_{\TRACE},\{\pi_i\}_{i\in I},n)=\sum_{k\in
K}p_k\cdot\PR_{\SCH_1}^{\mu_k}(C_{\TRACE},\EACT(\pi_k),n)$$ where
$\pi_k=\pi_i$ for any $s_i\in\SUPP(\mu_k)$.

  Now we prove by induction on $n$ that for each partial information
scheduler $\SCH_1$, there exists a partial information scheduler
$\SCH_2$, such that
  $$\PR_{\SCH_1}^{\mu}(C_{\TRACE},\{\pi_i\}_{i\in I},n)\le\PR_{\SCH_2}^{\nu}(C_{\TRACE},\{\pi'_i\}_{i\in J})$$
  for any $n\ge 0$ and $\TRACE$.

  First we assume that $\mu$ is transition consistent, which indicates
$\TC{\nu}$ by Lemma~\ref{lem:consistent preserved}. This is equivalent
to show that
  $$\PR_{\SCH_1}^{\mu}(C_{\TRACE},\EACT(\pi_i),n)\leq\PR_{\SCH_2}^{\nu}(C_{\TRACE},\EACT(\pi'_j)).$$
  We distinguish the following cases:
  \begin{enumerate}
  \item $n=0$ or $|\TRACE| = 0$. This case is trivial.
  \item $n > 0$, $|\TRACE| > 0$, and there exists
$\mu\TRANPsingle{\tau}\mu'=\sum_{k\in K}p_k\cdot\mu_k$ such that
$\TC{\mu_k}$ for each $k\in K$, and
    $$\PR_{\SCH_1}^{\mu}(C_{\TRACE},\EACT(\pi_i),n)  = \sum_{k\in K}p_k\cdot\PR_{\SCH_1}^{\mu_k}(C_{\TRACE},\EACT(\pi_i\circ(\tau,0,s_k)),n-1)$$
    for any $i\in I$ and $s_k\in\SUPP(\mu_k)$.  Suppose $|K|~=1$
i.e. $\TC{\mu'}$ and $\EACT(\mu')=\EACT(\mu)$, then
$\EACT(\pi_i\circ(\tau,0,s_k))=\EACT(\pi_i)$, thus
    $$\PR_{\SCH_1}^{\mu}(C_{\TRACE},\EACT(\pi_i),n)=\PR_{\SCH_1}^{\mu'}(C_{\TRACE},\EACT(\pi_i),n-1).$$
    Since $\mu~\LWBS~\nu$, there exists $\nu\WTRANP{\tau}\nu'$ such
that $\mu'~\LWBS~\nu'$, let $\SCH_2$ be a partial information
scheduler mimicking the transition of $\nu$, moreover by induction
    $$\PR_{\SCH_2}^{\nu'}(C_{\TRACE},\EACT(\pi_i))\ge\PR_{\SCH_1}^{\mu'}(C_{\TRACE},\EACT(\pi_i),n-1).$$

    If $|K|~=1$ and $\EACT(\mu')\neq\EACT(\mu)$, or $|K|~>1$, then
    $$\PR_{\SCH_1}^{\mu}(C_{\TRACE},\EACT(\pi_i),n)=\sum_{k\in K}p_k\cdot\PR_{\SCH_1}^{\mu_k}(C_{\TRACE},\EACT(\pi_i\circ(\tau,0,s_k)),n-1)$$ for any $s_k\in\SUPP(\mu_k)$.
    Since $\mu~\LWBS~\nu$, there
exists $$\nu\WTRANP{\tau}\nu'=\sum_{k\in K}p_k\cdot\nu_k$$ such that
$\mu_k~\LWBS~\nu_k$, thus $\TC{\nu_k}$ by Lemma~\ref{lem:consistent
preserved} for each $k\in K$, moreover $\EACT(\mu_k)=\EACT(\nu_k)$.
Let $\SCH_2$ be a scheduler mimicking the transition
$\nu\WTRANP{\tau}\nu'$.  According to Definition~\ref{def:pis} such
partial information scheduler $\SCH_2$ always exists, since only
$\tau$ transitions are involved.  Since $\mu_k~\LWBS~\nu_k$,
    $$\PR_{\SCH_2}^{\nu_k}(C_{\TRACE},\EACT(\pi_i\circ(\tau,0,r_k)))\ge \PR_{\SCH_1}^{\mu_k}(C_{\TRACE},\EACT(\pi_i\circ(\tau,0,s_k)),n-1)$$ by induction, where $r_k\in\SUPP(\nu_k)$ for each $k\in K$. Therefore
    $$\PR_{\SCH_2}^{\nu}(C_{\TRACE},\EACT(\pi_i))\ge \PR_{\SCH_1}^{\mu}(C_{\TRACE},\EACT(\pi_i),n).$$
  \item $n > 0$, $\TRACE=\alpha\TRACE'$, and there exists
$\mu\TRANPsingle{\alpha}\mu'=\sum_{k\in K}p_k\cdot\mu_k$ such that
$\TC{\mu_k}$ for each $k\in K$, and
    $$\PR_{\SCH_1}^{\mu}(C_{\TRACE},\EACT(\pi_i),n)  = \sum_{k\in K}p_k\cdot\PR_{\SCH_1}^{\mu_k}(C_{\TRACE'},\EACT(\pi_i\circ(\alpha,0,s_k)),n-1)$$
    for any $i\in I$ and $s_k\in\SUPP(\mu_k)$.  Since $\mu~\LWBS~\nu$,
there exists $\nu\WTRANP{\alpha}\sum_{k\in K}p_k\cdot\nu_k$ such that
$\mu_k~\LWBS~\nu_k$ for each $k\in K$. Let $\SCH_2$ be the scheduler
which mimic the weak transition of $\nu$. The $\SCH_2$ is guaranteed
to be a partial information scheduler, since all states will perform a
transition with label $\alpha$. By induction we have:
    $$\PR_{\SCH_2}^{\nu_k}(C_{\TRACE'},\EACT(\pi_i\circ(\alpha,0,r_k)))\ge \PR_{\SCH_1}^{\mu_k}(C_{\TRACE'},\EACT(\pi_i\circ(\alpha,0,s_k)),n-1)$$ where $r_k\in\SUPP(\nu_k)$ for each $k\in K$, therefore
    $$\PR_{\SCH_2}^{\nu}(C_{\TRACE},\EACT(\pi_i))\ge \PR_{\SCH_1}^{\mu}(C_{\TRACE},\EACT(\pi_i),n).$$
    The case when $\TRACE=\beta\TRACE'$ such that $\beta\neq\alpha$ is
trivial, since $\PR_{\SCH_1}^{\mu}(C_{\TRACE},\EACT(\pi_i),n)=0$.
  \item $n>0$, $n > 0$, $\TRACE=\lambda\TRACE'$. This case is similar
as Case 3, and is omitted here.
  \end{enumerate}

  Secondly, if $\neg\TC{\mu}$ and $\mu=\sum_{k\in K}\mu_k$ such that
$\TC{\mu_k}$ for each $k\in K$,
then $$\PR_{\SCH_1}^{\mu}(C_{\TRACE},\{\pi_i\}_{i\in I},n)=\sum_{k\in
K}p_k\cdot\PR_{\SCH_1}^{\mu_k}(C_{\TRACE},\EACT(\pi_k),n)$$ where
$\pi_k=\pi_i$ for any $s_i\in\SUPP(\mu_k)$.  Since $\mu~\LWBS~\nu$,
there exists $\nu\WTRANP{\tau}\sum_{k\in K}p_k\cdot\nu_k$ such that
$\mu_k~\LWBS~\nu_k$ for each $k\in K$.  Since $\TC{\mu_k}$ and we have
proved that there exists $\SCH_2$ such that
  $$\PR_{\SCH_2}^{\nu_k}(C_{\TRACE},\EACT(\pi'_k))\ge \PR_{\SCH_1}^{\mu_k}(C_{\TRACE},\EACT(\pi_k),n)$$
  for each $k\in K$, again let $\SCH_2$ mimic the transition of $\nu$
in a stepwise manner, we get
  $$\PR_{\SCH_2}^{\nu}(C_{\TRACE},\{\pi'_i\}_{i\in J})\ge\PR_{\SCH_1}^{\mu}(C_{\TRACE},\{\pi_i\}_{i\in I},n)$$ as desired.

  Note that $$\PR_{\SCH_1}^{\mu}(C_{\TRACE},\{\pi_i\}_{i\in
I})=\lim_{n\rightarrow\infty}\PR_{\SCH_1}^{\mu}(C_{\TRACE},\{\pi_i\}_{i\in
I},n),$$ the remaining proof is then straightforward. This completes
the proof.  
\end{enumerate}
\end{proof}

\subsection{Proof of Theorem~\ref{thm:distributed composition}}
\begin{proof}
  \begin{enumerate}
  \item In case that $s_1$ and $s_2$ contain no parallel operators,
    all schedulers of $s_1$ and $s_2$ are distributed schedulers according to
    Definition~\ref{def:distributed scheduler}. Therefore $s_1~\LWBS~s_2$
    implies $s_1~\LWBS_{\schedClassDistributedSchedulers}~s_2$ and vice versa.
  \item 
Let $\MC{R}=\{(\APAR{\mu_1}{\mu_3},\APAR{\mu_2}{\mu_3})\mid \mu_1~\LWBS_{\schedClassDistributedSchedulers}~\mu_2\}$,
  it suffices to prove that $\MC{R}$ is a late weak bisimulation with respect to $\schedClassDistributedSchedulers$.
  Let $(\APAR{\mu_1}{\mu_3})~\MC{R}~(\APAR{\mu_2}{\mu_3})$ and $\APAR{\mu_1}{\mu_3}\scheTran{\theta}{\SCH_1}\nu$ 
  for some $\SCH_1\in\schedClassDistributedSchedulers$, we shall show that there exists $\APAR{\mu_2}{\mu_3}\scheWTran{\theta}{\SCH_2}\nu'$
  for some $\SCH_2\in\schedClassDistributedSchedulers$ such that $\mu~\MC{R}~\nu$. We distinguish several cases:
  \begin{enumerate}
  \item $\theta\in\ACT$ and $\theta\not\in A$:\\
    Since $\SCH_1$ is a distributed scheduler, we have either
    \begin{inparaenum}[\scriptsize$(i)$]
      \item $\mu_1\scheTran{\theta}{\SCH_1}\nu_1$ such that $\nu=\APAR{\nu_1}{\mu_3}$, or
      \item $\mu_3\scheTran{\theta}{\SCH_1}\mu'_3$ such that $\nu=\APAR{\mu_1}{\mu'_3}$. 
    \end{inparaenum}
    We first consider Case $\scriptsize{i}$). Since $\mu_1~\LWBS_{\schedClassDistributedSchedulers}~\mu_2$,
    there exists $\mu_2\scheWTran{\theta}{\SCH_2}\nu_2$ for some $\SCH_2\in\schedClassDistributedSchedulers$ 
    such that $\nu_1~\LWBS_{\schedClassDistributedSchedulers}~\nu_2$,
    therefore there exists $\APAR{\mu_2}{\mu_3}\scheWTran{\theta}{\SCH_2}\APAR{\nu_2}{\mu_3}$.
    According to the definition of $\MC{R}$, we have $\nu=(\APAR{\nu_1}{\mu_3})~\MC{R}~(\APAR{\nu_2}{\mu_3})=\nu'$ as desired.
    The proof of Case $\scriptsize{ii}$) is similar and omitted here.
  \item $\theta\in A$:\\
    As before $\SCH_1$ is a distributed scheduler, according to the definition of parallel operator, it must be the case that
    $\mu_1\scheTran{\theta}{\SCH_1}\nu_1$ and $\mu_3\scheTran{\theta}{\SCH_1}\mu'_3$ such that $\nu=\APAR{\nu_1}{\mu'_3}$.
    Since $\mu_1~\LWBS_{\schedClassDistributedSchedulers}~\mu_2$, there exists $\mu_2\scheWTran{\theta}{\SCH_2}\nu_2$ such that
    $\nu_1~\LWBS_{\schedClassDistributedSchedulers}~\nu_2$, hence there exists $\APAR{\mu_2}{\mu_3}\scheWTran{\theta}{\SCH_2}\APAR{\nu_2}{\mu'_3}$
    such that $\nu=(\APAR{\nu_1}{\mu'_3})~\MC{R}~(\APAR{\nu_2}{\mu'_3})=\nu'$.
  \item $\theta=\lambda\in\RPLUS$:\\
    By the definition of parallel operator, we have
    $\mu_1\scheTran{\lambda_1}{\SCH_1}\nu_1$ and $\mu_3\scheTran{\lambda_2}{\SCH_1}\mu'_3$ such that $\lambda=\lambda_1+\lambda_2$ and
    $\nu=\frac{\lambda_1}{\lambda}\cdot(\APAR{\nu_1}{\mu_3})+\frac{\lambda_2}{\lambda}\cdot(\APAR{\mu_1}{\mu'_3})$. 
    Since $\mu_1~\LWBS_{\schedClassDistributedSchedulers}~\mu_2$, there exists $\mu_2\scheWTran{\lambda_1}{\SCH_2}\nu_2$
    such that $\nu_1~\LWBS_{\schedClassDistributedSchedulers}~\nu_2$, the remaining proof is straightforward
    based on the above proof.
  \end{enumerate}
\end{enumerate}
\end{proof}

In order to prove Theorem~\ref{thm:trace congruence}, we shall
introduce the following lemma:
\begin{lemma}\label{lem:partial and distributed}
  Let $\schedClass=\schedClassPartialInformation\cap\schedClassDistributedSchedulers$,
  then $\mu_1~\LWBS_{\schedClass}~\mu_2$ implies
  \begin{enumerate}
  \item $\mu_1~\TREQUIV[\schedClass]~\mu_2$;
  \item $\APAR{\mu_1}{\mu_3}~\LWBS_{\schedClass}~\APAR{\mu_2}{\mu_3}$ for any $\mu_3$.
  \end{enumerate}
\end{lemma}
\begin{proof}
  \begin{enumerate}
  \item Refer to the proof of Theorem~\ref{thm:implication}.
  \item The proof is similar as the proof of Clause 2 of Theorem~\ref{thm:distributed composition}.
  \end{enumerate}
\end{proof}

\subsection{Proof of Theorem~\ref{thm:trace congruence}}
\begin{proof}
  \begin{itemize}
  \item $\LWBS_{\schedClass}~\Rightarrow~\traceCongruence$:\\
    \begin{align*}
      \mu_1~\LWBS_{\schedClass}~\mu_2  &\overset{Lem.~\ref{lem:partial and distributed}}{\Longrightarrow}
      \mu_1~\TREQUIV[\schedClass]~\mu_2\text{ and }\APAR{\mu_1}{\mu_3}~\LWBS_{\schedClass}~\APAR{\mu_2}{\mu_3}\\
      &\overset{\text{Def. of }\traceCongruence}{\Longrightarrow}\mu_1~\traceCongruence~\mu_2                  
    \end{align*}
  \item $\traceCongruence~\Rightarrow~\LWBS_{\schedClass}$:\\
    Let $\MC{R}=\{(\mu_1,\mu_2)\mid\mu_1~\traceCongruence~\mu_2\}$, we show that $\MC{R}$ is a late weak 
    bisimulation with respect to $\schedClass$. Let $\mu_1~\MC{R}~\mu_2$. We first assume that $\TC{\mu_1}$ and
    $\mu_1\scheTran{\theta}{\SCH_1}\mu'_1$ for some $\theta$ and $\SCH_1\in\schedClass$. We need to prove
    that there exists $\mu_2\scheWTran{\theta}{\SCH_2}\mu'_2$ for some $\SCH_2\in\schedClass$ such that
    $\mu'_1~\MC{R}~\mu'_2$. We proceed by contradiction and assume that $\mu'_1~\not\MC{R}~\mu'_2$ i.e.
    $\mu'_1~\not\traceCongruence~\mu'_2$, we distinguish several cases as follows, where the main
    idea is to construct a distribution $\mu_3$ with a proper set $A$ such that 
    $\APAR{\mu_1}{\mu_3}~\not\TREQUIV[\schedClass]~\APAR{\mu_2}{\mu_3}$.
    \begin{enumerate}
    \item $\theta\in\ACT$ and $\mu'_1~\not\TREQUIV[\schedClass]~\mu'_2$:\\
      Given a set of visible actions $A$, we let
      $s'=A.s'$ denote a state which can only perform self loop transitions with labels in $A$.
      We can see that for any distribution $\mu$ such that $\TC{\mu}$,
      $\APAR{\mu}{\DIRAC{s'}}$ induces the same trace distribution as $\mu$, where
      $A$ contains all possible actions which can be performed by states in $\SUPP(\mu)$ and their successors.
      
      Now let $A$ contains all visible actions which can be performed by states in $\SUPP(\mu_1)$ and $\SUPP(\mu_2)$
      and their successors. Let $s=\theta.s'$ where $s'$ is defined as above. Then for each $\SCH_2\in\schedClass$,
      there exists $\SCH_1\in\schedClass$
      such that $\PR_{\APAR{\mu_1}{\DIRAC{s}}}^{\SCH_1}(C_{\theta\TRACE})=\PR_{\APAR{\mu'_1}{\DIRAC{s'}}}^{\SCH_2}(C_{\TRACE})=\PR_{\mu'_1}^{\SCH_2}(C_{\TRACE})$,
      for each $\TRACE$, similarly for $\APAR{\mu_2}{\DIRAC{s}}$. Since $\mu'_1~\not\TREQUIV[\schedClass]~\mu'_2$, 
      we conclude that $\APAR{\mu_1}{\DIRAC{s}}~\TREQUIV[\schedClass]~\APAR{\mu_2}{\DIRAC{s}}$, which
      contradicts the assumption that $\mu_1~\traceCongruence~\mu_2$ ($\mu_3=\DIRAC{s}$).
    \item $\theta\in\ACT$ and $\mu'_1~\TREQUIV[\schedClass]~\mu'_2$, but there exists $\mu'_3$ and $A'$ 
      such that $\APAR[A']{\mu'_1}{\mu'_3}~\not\TREQUIV[\schedClass]~\APAR[A']{\mu'_2}{\mu'_3}$:\\
      If $\theta\in A'$, we can simply let $\mu_3=\theta.\mu'_3$ and $A=A'$ i.e. the only immediate transition of $\mu_3$ 
      is $\mu_3\TRAN{\theta}\mu'_3$, and the remaining argument is similar as Case 1.
      
      Now suppose that $\theta\not\in A$. For each $s\in\SUPP(\mu'_3)$, we let $s_\theta$ denote the copy
      of $s$ but by adding a self loop with label $\theta$ to $s$ and all its successors. Let $A=A'\cup\{\theta\}$ and
      $\mu_3$ be a distribution such that the only immediate transition of $\mu_3$ is
      $\mu_3\TRAN{\theta}\mu''_3$ where $\mu''_3(s_\theta)=\mu'_3(s)$ for each $s\in\SUPP(\mu'_3)$,
      then for each $\SCH_1\in\schedClass$, there exists $\SCH_2\in\schedClass$ such that
      $\PR^{\SCH_1}_{\APAR[A']{\mu'_1}{\mu'_3}}(C_\TRACE)=\PR^{\SCH_2}_{\APAR{\mu'_1}{\mu''_3}}(C_\TRACE)$ for each $\TRACE$, similarly
      for $\APAR{\mu'_2}{\mu'_3}$ and $\APAR{\mu'_2}{\mu''_3}$.
      Therefore $(\APAR{\mu'_1}{\mu''_3})~\not\TREQUIV[\schedClass]~(\APAR{\mu'_2}{\mu''_3})$. Since $\theta\in A$, 
      the remaining proof is similar.
    \item $\theta\in\RPLUS$:\\
      Since $\mu'_1~\not\MC{R}~\mu'_2$, we have either i) $\mu'_1~\not\TREQUIV[\schedClass]~\mu'_2$, or ii) there exists 
      $\mu'_3$ and $A'$ such that $\APAR[A']{\mu'_1}{\mu'_3}~\not\TREQUIV[\schedClass]~\APAR[A']{\mu'_2}{\mu'_3}$.
      For Case i) let $A=A'$, $s'$ be defined as above, and $\mu_3=\DIRAC{\alpha.s'}$ such that $\alpha$ is a fresh action.
      Then for each $\SCH_2\in\schedClass$, there exists $\SCH_1\in\schedClass$ such that
      $\PR_{\APAR{\mu'_1}{\mu_3}}^{\SCH_1}(C_{\alpha\TRACE})=\PR_{\mu'_1}^{\SCH_2}(C_\TRACE)$ for each $\TRACE$,
      therefore $\APAR{\mu'_1}{\mu_3}~\not\TREQUIV[\schedClass]~\APAR{\mu'_2}{\mu_3}$. Since there exists only one 
      transition labelled with $\theta\in\RPLUS$, hence
      for each $\SCH_2\in\schedClass$, there exists $\SCH_1\in\schedClass$ such that 
      $\PR_{\APAR{\mu_1}{\mu_3}}^{\SCH_1}(C_{\theta\TRACE})=\PR_{\APAR{\mu'_1}{\mu_3}}^{\SCH_2}(C_\TRACE)$ for each $\TRACE$, similarly for $\APAR{\mu_2}{\mu_3}$.
      Therefore we conclude that $\APAR{\mu_1}{\mu_3}~\not\TREQUIV[\schedClass]~\APAR{\mu_2}{\mu_3}$, which
      contradicts the assumption that $\mu_1~\traceCongruence~\mu_2$.
      
      For Case ii), we can let $A=A'$ and $\mu_3=\alpha.\mu'_3$ i.e. the only immediate transition of $\mu_3$ is
      $\mu_3\TRAN{\alpha}\mu'_3$ where $\alpha$ is a fresh action. The remaining argument is similar as Case i).
    \end{enumerate}
    For now we have only considered case when $\mu_1$ and $\mu_2$ are transition consistent.  In case that $\mu_1$ is not transition consistent, we can always find a split $\mu=\sum_{i\in I}p_i\cdot\nu_i$
    such that $\sum_{i\in I}p_i=1$ and $\TC{\nu_i}$ for each $i\in I$, moreover there exists
    $\mu_2\scheWTran{\tau}{\SCH}\sum_{i\in I}p_i\cdot\nu'_i$ such that $\nu_i~\traceCongruence~\nu'_i$ for each $i\in I$.
    Then we can apply the same arguments as when $\mu_1$ is transition consistent.
  \end{itemize}
\end{proof}

\end{document}

\section{Response to the reviews}
Q1: Does $s\TRANP{\tau}\WTRAN{\theta,n-1}\mu_1$ and $s\TRANP{\theta}\WTRAN{\tau,n-1}\mu_2$
imply $s\WTRANP{\theta}\mu\equiv(\frac{1}{2}\cdot\mu_1 + \frac{1}{2}\cdot\mu_2)$?

Answer: Yes. This is guaranteed by the definition of combined transitions (page 6, line -8). Since 
the combined transitions is closed under  finite linear combinations. Maybe it will be clearer if we define 
$\WTRAN{\theta, n}$ instead of $\WTRANP{\theta, n}$ in the proof of Lemma 1. 

Q2: It is not obvious the limit of $\mu'_n$ is $\mu'$.

Answer: We have further improved the proof of Lemma 1 with more details. Basically, the idea is to show
the step-bounded weak transition is able to mimic the weak transition defined in page 6 in a stepwise manner.
Therefore both of them converge to the same limit. 

Here
$$\PR_{\SCH_1}^{\mu}(C_{\TRACE},\{\pi_i\}_{i\in I})=\sum_{i\in I}\PR_{\SCH_1}^{s_i}(C_{\TRACE},\pi_i),$$
and similarly for $\PR_{\SCH_2}^{\nu}(C_{\TRACE},\{\pi'_i\}_{i\in
J})$.  Moreover $\PR_{\SCH_1}^{s_i}$ can be defined as follows where
$\TRACE=\alpha\TRACE'$: $\PR_{\SCH_1}^{s_i}(C_{\TRACE},\pi) = $
$$
\begin{aligned}
&\sum_{(\alpha,\mu')\in\SUPP(\SCH_1(\pi_i,\cdot))}\SCH(\pi_i,\alpha,\mu')\cdot\sum_{s'\in\SUPP(\mu')}\mu'(s')\cdot\PR_{\SCH_1}^{s'}(C_{\TRACE'},\pi_i\circ(\alpha,0,s'))\\
+ &
\sum_{(\tau,\mu')\in\SUPP(\SCH(\pi_i,\cdot))}\SCH(\pi_i,\tau,\mu')\cdot\sum_{s'\in\SUPP(\mu')}\mu'(s')\cdot\PR_{\SCH_1}^{s'}(C_{\TRACE},\pi_i\circ(\tau,0,s')).
\end{aligned}
$$
Since $\SCH_1$ is a partial information scheduler, we can rewrite the
above equation as follows by overloading $\SCH_1$ in a straightforward
way: $\PR_{\SCH_1}^{s_i}(C_{\TRACE},\EACT(\pi_i)) = $
$$
\begin{aligned}
&\sum_{(\alpha,\mu')\in\SUPP(\SCH_1(\EACT(\pi_i),\cdot))}\SCH_1(\EACT(\pi_i),\alpha,\mu')\cdot\sum_{s'\in\SUPP(\mu')}\mu'(s')\cdot\PR_{\SCH_1}^{s'}(C_{\TRACE'},\EACT(\pi_i\circ(\alpha,0,s'))\\
+ &
\sum_{(\tau,\mu')\in\SUPP(\SCH_1(\EACT(\pi_i),\cdot))}\SCH_1(\EACT(\pi_i),\tau,\mu')\cdot\sum_{s'\in\SUPP(\mu')}\mu'(s')\cdot\PR_{\SCH_1}^{s'}(C_{\TRACE},\EACT(\pi_i\circ(\tau,0,s'))).
\end{aligned}
$$

